\documentclass[journal]{IEEEtran}[12pt]
\ifCLASSINFOpdf
  \usepackage[pdftex]{graphicx}
  \graphicspath{{../pdf/}{../jpeg/}}
  \DeclareGraphicsExtensions{.pdf,.jpeg,.png}
\else
  \usepackage[dvips]{graphicx}
  \graphicspath{{../eps/}}
  \DeclareGraphicsExtensions{.eps}
\fi\usepackage{graphicx}

\usepackage{graphics}
\usepackage{epsfig}
\usepackage{epstopdf}
\usepackage{stfloats}
\usepackage[cmex10]{amsmath}
\usepackage{algorithmic}
\usepackage{array}
\usepackage{mdwmath}
\usepackage{mdwtab}
\usepackage{graphicx}
\usepackage{subfigure}
\usepackage{color}
\usepackage{amsfonts,amssymb}
\usepackage{hyperref} 
\usepackage{multirow}
\usepackage{diagbox}
\usepackage{caption}
\usepackage{verbatim}
\usepackage{float}
\usepackage{graphicx}
\usepackage{bm}

\usepackage{amsfonts,amsthm,array} 
\makeatletter
\renewcommand{\maketag@@@}[1]{\hbox{\m@th\normalsize\normalfont#1}}
\makeatother

\newtheorem{remark}{Remark}

\begin{document}

\title{Dynamic Resource Management in CDRT Systems through Adaptive NOMA \thanks{Manuscript received.}}

\author{Hongjiang~Lei, 
	Mingxu~Yang,
	Ki-Hong~Park, 
	Nasir Saeed, \\
	Xusheng~She,
	and~Jianling~Cao
	\thanks{This work was supported by the National Natural Science Foundation of China under Grant 61971080. Corresponding author: \textit{Jianling~Cao}.}
	\thanks{H. Lei, M. Yang, X. She, and J. Cao are with the School of Communication and Information Engineering, Chongqing University of Posts and Telecommunications, Chongqing 400065, China (e-mail: leihj@cqupt.edu.cn, ymx13648245248@163.com, sxs17300289459@163.com, caojl@cqupt.edu.cn).}
	\thanks{K.-H. Park is with CEMSE Division, King Abdullah University of Science and Technology (KAUST), Thuwal 23955-6900, Saudi Arabia (e-mail: kihong.park@kaust.edu.sa).}
	\thanks{N. Saeed is with the Department of Electrical and Communication Engineering, United Arab Emirates University (UAEU), Al Ain, United Arab Emirates (e-mail: mr.nasir.saeed@ieee.org).}
}

\maketitle
\begin{abstract}
This paper introduces a novel adaptive transmission scheme to amplify the prowess of coordinated direct and relay transmission (CDRT) systems rooted in non-orthogonal multiple access principles. 
Leveraging the maximum ratio transmission scheme, we seamlessly meet the prerequisites of CDRT while harnessing the potential of dynamic power allocation and directional antennas to elevate the system's operational efficiency. 
Through meticulous derivations, we unveil closed-form expressions depicting the exact effective sum throughput. 
Our simulation results adeptly validate the theoretical analysis and vividly showcase the effectiveness of the proposed scheme.
\end{abstract}


\section{Introduction}
\label{sec:introduction}

\subsection{Background and Related Work}
In the era of the Internet of Things (IoT), addressing the pressing challenge of managing scarce spectrum resources while facilitating seamless connectivity for a multitude of devices has emerged as a critical concern. Among the various techniques, non-orthogonal multiple access (NOMA) has surfaced as a promising solution. This approach leverages superimposed coding at the transmitter and employs successive interference cancellation (SIC) at the receiver, effectively catering to distinct rate and delay requirements \cite{DingZ2017JSAC}.
Cooperative NOMA (CNOMA) technology, proven and prolific, stands out for its capacity to significantly extend network coverage and elevate overall system performance \cite{LvL2018Mag, WanD2018WC, ZengM2020Net}. Building on this foundation, Kim \emph{et al.} explored CNOMA systems, yielding incisive closed-form expressions for the ergodic sum rate (ESR) \cite{KimJ2015CL}. Their analytical revelations underscored the superior spectral efficiency (SE) intrinsic to CNOMA systems equipped with a dedicated relay, positioning them ahead of conventional cooperative systems.
Furthermore, Luo \emph{et al.} delved into the realm of CNOMA systems augmented by a buffer-assisted relay \cite{LuoS2017CL}. Their  work introduced an adaptive transmission scheme designed to maximize the cumulative throughput. In this pursuit, they derived expressions encapsulating the essence of the effective sum throughput (EST), unraveling new vistas for performance optimization.
Moreover, the work in \cite{MukherjeeA2022TWC} introduced underlay CNOMA paradigm. At its core lies in the empowered capability of a full-duplex near-user to serve as a conduit for relaying signals to a distant user. Herein, adaptive strategies, spanning cooperative and non-cooperative modes along with transmit antenna selection, were proposed to get better performance. Deftly, the authors derived expressions illuminating the outage probability (OP) and the contours of the EST, further underlining the potency of this innovative approach.

Additionally, multi-antenna technology can significantly increase the CNOMA system's performance due to beamforming array gain \cite{LiuYW2018WC}. In this context, \cite{MenJ2015CL} navigated through the intricacies of a CNOMA configuration embellished with a multi-antenna relay. Notably, the study encompassed the derivation of closed-form expressions for both the exact and lower bound of the OP, tailored explicitly for scenarios where the relay's transmit antenna was thoughtfully selected.
Meanwhile, Han \emph{et al.} delved into multi-antenna satellite cooperative systems, investigating fixed and variable gain relay schemes, all within the backdrop of imperfect SIC \cite{HanL2021CL}. They derived analytical expressions encapsulating the exact and asymptotic OP.
Moreover, Lv \emph{et al.} investigated the potential of cooperative NOMA domain infused with multi-antenna two-way relays. Notably, their work yielded two novel cooperative schemes: multiple-access NOMA and time-division NOMA \cite{LvL2020TWC}. Their work involved ingenious transmission paradigms, wherein antenna and relay were inextricably interwoven through joint selection. Importantly, this holistic study culminated in formulating expressions that elegantly encapsulated both the exact and asymptotic OP, adding further depth to the understanding of these configurations.

NOMA-based coordinated direct and relay transmission (CDRT) has garnered substantial attention as a promising strategy to bolster system capacity \cite{KimJB2015CL}-\cite{AnandJ2022IoT}. Notably distinct from CNOMA, the CDRT scheme capitalizes on the full range of SIC results at the near-user. This attribute effectively curtails interference from the relay to the near-user during the second time slot, thus augmenting the efficiency of parallel link transmission. This configuration efficiently enhances system SE by transmitting multiple signals to NOMA users on the same resource block.
An early stride into this work was taken with the introduction of a downlink CDRT system \cite{KimJB2015CL}, where analytical expressions for the OP and ESR were elucidated. Notably, this work showcased that CDRT systems' ESR surpasses conventional CNOMA systems.
A subsequent investigation by Liu \emph{et al.} explored the outage performance of a satellite-based CDRT system, culminating in closed-form expressions capturing the exact and asymptotic OP \cite{LiX2020CL}. Meanwhile, Zou \emph{et al.} ventured into device-to-device CDRT systems, deriving the closed-form ESR expression and revealing its dependence on the relay-near user distance and power allocation \cite{ZouL2020CL}.
The exploration further intensified with a dynamic transmission scheme proposed by Xu \emph{et al.}, where near-users alternated between forward and receive modes based on the first slot decoding outcomes \cite{XuY2018IET}. Their insights led to closed-form expressions encompassing exact and asymptotic OP and ESR, while power allocation coefficients were meticulously optimized.
Towards a holistic design, Xu \emph{et al.} introduced a physical layer network coding-infused CDRT scheme, fostering joint uplink and downlink transmission advancements \cite{XuY2021TWC}. The outcome was the derivation of closed-form expressions capturing the OP, EST, and ESR for scenarios embracing perfect and imperfect SIC.
Shifting focus towards finite block lengths, Yuan \emph{et al.} delved into CDRT system performance, yielding an approximate expression for the EST alongside an optimized power allocation coefficient \cite{YuanL2022WCL}.
With direct links, amplify-and-forward (AF), and decode-and-forward (DF) relays, Anand \emph{et al.} embarked on an intricate investigation \cite{AnandJ2022TCOM}. Their findings culminated in closed-form expressions for exact and asymptotic OP and EST, orchestrating an optimization interplay involving power allocation coefficients and rate thresholds to achieve maximal EST while preserving far-user quality of service.
In incremental AF relay-assisted CDRT systems, Anand \emph{et al.} traversed the domain of imperfect SIC \cite{AnandJ2022IoT}. Their findings underscored the potency of incremental signaling and dual combining at the far-user and near-user, harmoniously enriching throughput and energy efficiency.

\subsection{Motivation and Contributions}
Upon delving into the array of existing studies, it becomes apparent that the performance of CDRT systems has been thoroughly examined across various scenarios. However, a conspicuous gap remains in the realm of investigating beamforming techniques and dynamic power allocation (DPA) strategies within the CDRT framework, serving as the impetus for the current paper. It's important to underscore the foundational premise of the CDRT system, which hinges upon the successful decoding of signals by the near-user for the edge-user \cite{LeiH2023CDRT}. If this pivotal condition is not met, the relay's interference can hinder the parallel transmission's effectiveness. Therefore, ensuring the near users' adept decoding of edge-user signals is a critical challenge within the CDRT framework.
In this study, we propose an innovative adaptive NOMA-based CDRT scheme that dynamically adjusts the transmit power at the transmitter to guarantee the relay's successful decoding of signals intended for the edge-user. This strategy is underpinned by the rationale that, when close to the relay, the near-user's proximity to the transmitter assures the relay's accurate decoding of edge-user information. Consequently, the near-user exhibits an elevated likelihood of successful signal decoding, underpinning the rationale for this approach. Notably, this scheme can improve edge-user performance, given that the edge-user's signal-to-noise ratio (SNR) is contingent on the SNRs of both transmission hops.
Moreover, to further enrich the near-user experience quality, we harness a beamforming scheme that optimally directs signal propagation. Simultaneously, we employ directional antenna transmission to mitigate interference at the near user's end.
The main contributions of th are summarized as follows.

\begin{enumerate}
	\item An adaptive CDRT scheme is proposed to enhance performance through DPA and beamforming schemes. More specifically, the maximum ratio transmission (MRT) scheme is utilized, and the transmitter sends superimposed signals with a DPA scheme to ensure that the relay can successfully decode the signals for the edge-user. Meanwhile, a beamforming strategy is adopted to increase the probability that center-user will successfully decode the target signal.
	
	\item The closed-form expressions of exact EST for the proposed scheme are derived. To attain more insights, we adopt the single/multiple antenna based-CDRT schemes with fixed power allocation (FPA) and beamforming-CDRT scheme between transmitter and relay to enhance the quality of edge-user's information as a benchmark to prove that the proposed scheme can achieve superior performance in EST. 	Then, the effect of parameters such as the distance between the transmitter and near user/the relay and the rate threshold on EST was analyzed. Monte Carlo simulation results are provided to prove the accuracy of the derived analytical expressions.
	
	\item Relative to \cite{YuanL2022WCL} wherein different scenarios in which the relay can or cannot decode the signals for the edge-user were considered. However, the condition for successful decoding was not considered and no schemes were proposed to deal with these events.	The power allocation coefficient is designed to ensure that the relay successfully decodes the information for the edge-user, and beamforming technology is utilized to improve system performance. Moreover, the closed-form expressions of exact EST are derived.
	
\item Relative to \cite{VuT2022TVT} wherein the transmitter utilizes beamforming technology to improve the performance of the downlink NOMA system, the closed-form expressions for the approximate and asymptotic block error rate and ESR are derived. We study the downlink CDRT system with a DPA strategy, and closed-form expressions of exact throughput OP and EST are derived. Technically speaking, it is much more challenging to study the performance of the CDRT system than the NOMA system.
	
\end{enumerate}

\subsection{Organization}
The rest of this work is organized as follows.
Section \ref{sec:SystemModel} describes the system model.
The EST of the considered systems is analyzed in Section \ref{ESTAnalysis}.
Section \ref{sec: RESULTS} presents the numerical and simulation results, and this work is concluded in Section \ref{sec: Conclusion}.

\section{System Model}
\label{sec:SystemModel}

\begin{figure*}[!t]
	\centering
	\includegraphics[width = 5 in]{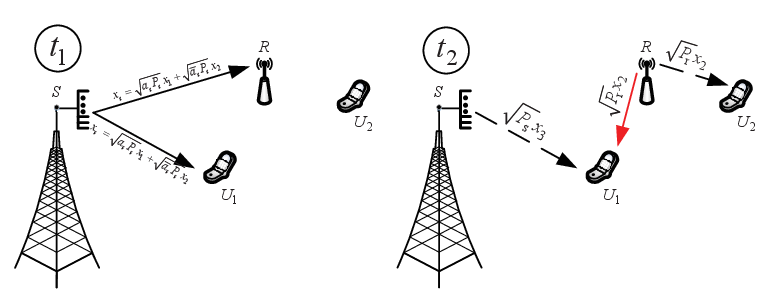}
	\caption{A NOMA-based CDRT system consisting of a transmitter ($S$), two users (${U_1}$ and ${U_2}$), and a relay (${R}$).}
	\label{figmodel}
\end{figure*}

Fig. \ref{figmodel} illustrates a NOMA-based CDRT system consisting of a transmitter $\left( S \right)$, a center-user $\left( {{U_1}} \right) $, and an edge-user $\left( {{U_2}} \right)$, where
all the nodes are equipped with a single antenna unless otherwise stated.
There is no direct link between $S$ and $U_2$ due to deep fading and shadowing, and the communication link between $S$ and $U_2$ must be deployed via a DF relay $\left( R \right)$.
Moreover, all the wireless links are assumed to experience quasi-static independent Rayleigh fading, and all nodes operate in the half-duplex mode.
To simplify the analysis, subscripts $s$, $r$, 1, and 2 denote the $S$, $R$, $U_1$, and $U_2$, respectively.
The channel coefficient and average channel gain between $p$ and $q$ are denoted by ${h_{p,q}}$ and ${\lambda _{p,q}} = d_{p,q}^{ - \alpha }$ for $p,q \in \left\{ {\rm{s}, 1, 2, r} \right\}$ $\left( {p \ne q} \right)$, where
${d_{p,q}}$ denotes the distance between $p$ and $q$ and $\alpha$ signifies the path loss exponent, respectively.
Furthermore, the distances are assumed to be ${d_{{\rm{s}},1}} < {d_{{\rm{s}},{\rm{r}}}} < {d_{{\rm{s}},2}}$ \cite{KimJB2015CL}.

The transmission block is divided into two equal time slots.
In the first time slot $\left( {{t_1}} \right)$,
${S}$ broadcasts a superimposed signal, ${x_{\rm{s}}} = \sqrt {{{{a  _1}}}{{P_{\rm{s}}}}} {x_1} + \sqrt {{{a  _2}}{{P_{\rm{s}}}}} {x_2}$,
where $x_i$ and ${a  _i}$ denote the signal and power allocation coefficient for $U_i$, respectively, $i = 1,2$,
${a _1} + {a  _2} = 1$,
and
${{P_{\rm{s}}}}$ denotes the transmit power at $S$.
In the second time slot $\left( {{t_2}} \right)$, $R$ decodes and forwards $x_2$ with power $P_{\rm{r}}$. Simultaneously, $S$ transmits a new signal ${x_3}$ to ${U_1}$.

\subsection{Dynamic Power when $S$ equipped with a Single Antenna}

The received signal at $U_1$ is expressed as
\begin{equation}
	y_1^{{t_1}} = {h_{{\rm{s1}}}}{x_{\rm{s}}} + n_1^{{t_1}},
	\label{yt1}
\end{equation}
where ${n_1^{{t_1}}}$ signifies the additive white Gaussian noise (AWGN) with zero mean and variance ${\sigma ^2}$.
Subsequently, ${U_1}$ utilizes SIC detection following the decoding order of ${{x_2} \to {x_1}}$ to obtain better performance on $x_1$ and the corresponding achievable rate of ${x_2}$ is expressed as
\begin{equation}
	\begin{aligned}
		R_1^{{x_2}} = \frac{1}{2}\ln \left( {1 + \gamma _1^{{x_2}}} \right),
		\label{rateU1x2}
	\end{aligned}
\end{equation}
where ${\gamma _1^{{x_2}} = \frac{{{\rho _{\rm{s}}}{a_2}{Y_1}}}{{1 + {\rho _{\rm{s}}}{a_1}{Y_1}}}}$,
${Y_1} = {\left| {{h_{{\rm{s}},1}}} \right|^2}$,
and
${\rho _{\rm{s}}} = \frac{{P_{\rm{s}}}}{\sigma ^2}$ denotes the normalized power at $S$.
Thus, the corresponding achievable rate of ${x_2}$ is expressed as
\begin{equation}
	\begin{aligned}
			R_1^{{x_1}} = \frac{1}{2}\ln \left( {1 + \gamma _1^{{x_1}}} \right),\,\,{\rm{when}} \,\,\,R_1^{{x_2}} \ge R_{{\rm{th}}}^{{x_2}},
		\label{rateU1x1}
	\end{aligned}
\end{equation}
where ${\gamma _1^{{x_1}} = {\rho _{\rm{s}}}{a_1}{Y_1}}$ and
$R_{{\rm{th}}}^{{x_2}}$ signifies the target rate threshold for $x_2$.

Meanwhile, $x_2$ is directly decoded at $R$ and the achievable rate is expressed as
\begin{equation}
	\begin{aligned}
		R_{\rm{r}}^{{x_2}} = \frac{1}{2}\ln \left( {1 + \gamma _{\rm{r}}^{{x_2}}} \right),
		\label{snrRx2}
	\end{aligned}
\end{equation}
where ${\gamma _{\rm{r}}^{{x_2}} = \frac{{{\rho _{\rm{s}}}{a_2}{X_1}}}{{1 + {\rho _{\rm{s}}}{a_1}{X_1}}}}$ and
${X_1} = {\left| {{h_{{\rm{s}},{\rm{r}}}}} \right|^2}$.

In the second time slot, $R$ decodes and forwards $x_2$ to $U_2$ and $S$ transmit $x_3$ to $U_1$.
The received signals at ${U_1}$ and $U_2$  are expressed as
\begin{equation}
	y_1^{{t_2}} = {{{h}}_{{\rm{s}},1}}\sqrt {{{P_{\rm{s}}}}} {x_3} + {h_{{\rm{r}},1}}\sqrt {{P_{\rm{r}}}} {x_2} + n_1^{{t_2}},
	\label{yU1t2}
\end{equation}
\begin{equation}
	y_2^{{t_2}} = {h_{{\rm{r}},2}}\sqrt {{P_{\rm{r}}}} {x_2} + n_2^{{t_2}},
	\label{yU2t2}
\end{equation}
respectively,
where
$n_1^{{t_2}}$ and $n_2^{{t_2}}$ denote the AWGN at $U_1$ and $U_2$ in $t_2$.
Since $x_2$ is decoded in the first slot at $U_1$ and can be deleted from $y_1^{{t_2}}$, then the achievable rate  of $x_3$ at $U_1$ is expressed as
\begin{equation}
	R_1^{{x_3}} = \frac{1}{2}\ln \left( {1 + \gamma_1^{{x_3}}} \right), \,\,{\rm{when}}\,R_1^{{x_2}} \ge R_{{\rm{th}}}^{{x_2}},
	\label{rateU1x3}
\end{equation}
where $\gamma _1^{{x_3}} = {\rho _{\rm{s}}}{Y_1}$.
And the achievable rate  of $x_2$ at $U_2$ is expressed as
\begin{equation}
	R_2^{{x_2}} = \frac{1}{2}\ln \left( {1 + \gamma_2^{{x_2}}} \right), \,\,{\rm{when}}\,R_{\rm{r}}^{{x_2}} \ge R_{{\rm{th}}}^{{x_2}},
	\label{rateU2x2}
\end{equation}
where
$\gamma_2^{{x_2}} = {\rho _{\rm{r}}}{\left| {{h_{{\rm{r}},2}}} \right|^2}$
and
${\rho _{\rm{r}}} = \frac{{P_{\rm{r}}}}{\sigma ^2}$ denotes the normalized power at $R$.

\subsubsection{Dynamic power based on $S$-$U_1$ link (DPU) in the first time slot}

As stated in \cite{LeiH2023CDRT}, there is a premise for the CDRT system that $U_1$ should successfully decode $x_2$ to eliminate the interference in the second slot.
Then the power allocation at $S$ is utilized to meet $R_1^{{x_2}} \ge R_{{\rm{th}}}^{{x_2}}$.
Based on (\ref{rateU1x2}), the condition is obtained as
\begin{equation}
	{a_1} \le \theta \left( {1 - \frac{{{\tau _{\rm{1}}}}}{{{Y_1}}}} \right),{Y_1} > {\tau _1},
	\label{a1constraint1}
\end{equation}
where $\theta  = \frac{1}{{1 + {\theta _2}}}$,
${\theta _2} = \exp \left( {2R_{{\rm{th}}}^{{x_2}}} \right) - 1$,
and
${\tau _1} = \frac{{{\theta _2}}}{{{\rho _{\rm{s}}}}}$.

\begin{remark}
	Based on (\ref{a1constraint1}), one can observe that ${Y_1} > {\tau _1}$ must be satisfied in this scenario to utilize the CDRT scheme.
	This signifies that $x_2$ can be successfully decoded when $x_2$ occupies the channel alone and all the power at $S$ is allocated to $x_2$. This is easy to follow since the NOMA scheme can improve the spectrum efficiency when the channel quality is relatively high.
\end{remark}

When ${Y_1} < {\tau _1}$, the system cannot work normally.

\subsubsection{Dynamic power based on $S$-$R$ link (DPR) in the first slot}

Only if $R$ is guaranteed to successfully decode $x_2$ to forward the information in the second time slot.
Another DPA scheme is proposed where the power allocation at $S$ is dynamically adjusted to meet
$R_{\rm{r}}^{{x_2}} \ge R_{{\rm{th}}}^{{x_2}}$.
Then the following condition is obtained as
\begin{equation}
	{a_1} \le \theta \left( {1 - \frac{{{\tau _{\rm{1}}}}}{{{X_1}}}} \right),{X_1} > {\tau _1}.
	\label{a1constraint2}
\end{equation}

For the scenarios with ${X_1} < {\tau _1}$, the system also cannot work normally. 

\subsection{Dynamic Power when $S$ equipped with Multiple Antennas}

Based on (\ref{a1constraint1}) and (\ref{a1constraint2}), one can observe that when ${Y_1} < {\tau _1}$ or ${X_1} < {\tau _1}$ are not satisfied, it is difficult to design power allocation coefficient to ensure that $U_1$ or $R$ are successfully decoded $x_2$.
Thus, a new adaptive transmission scheme, MDPR, is proposed in this subsection. 
We adopt $S$ equipped with $N$ antennas to address the above problem.
The MRT scheme is utilized on $S-{U_1}$ link to enhance the channel quality and the power at $S$ is dynamically adjusted based on $S$-$R$ link to meet $R_{\rm{r}}^{{x_2}} \ge R_{{\rm{th}}}^{{x_2}}$
\footnote{
	Compared with $R$, $U_1$ is assumed to be closer to $S$. The channel quality of the $S$-$U_1$ link is more robust than that of the $S$-$R$ link with high probability. When $R$ is guaranteed to decode $x_2$, $U_1$ can decode $x_2$ with a higher probability.
}
, and the corresponding received signals are expressed as
\begin{equation}
	y_1^{{t_1}} = {{\bf{h}}_{{\rm{s}},1}}{\bf{w}}{x_{\rm{s}}} + n_1^{{t_1}},
	\label{yt1ma}
\end{equation}
where
${{\bf{h}}_{{\rm{s}},1}}$ denotes the channel coefficients between the $S$ and $U_1$
and
$\mathbf w  = \frac{{{{\mathbf{h}}}_{{\rm{s}},1}^H}}{{\left\| {{{\mathbf{h}}_{{\rm{s}},1}}} \right\|}}$ is the the beamforming vector.
The corresponding achievable rate of ${x_i}$ $\left( {i = 1,2} \right)$ are expressed as
\begin{equation}
	R_{1,{\rm{noma}}}^{{x_2},\rm{m}} = \frac{1}{2}\ln \left( {1 + \gamma _{1,{\rm{noma}}}^{{x_2}}} \right),
	\label{snrU1x2ma}
\end{equation}
\begin{equation}
	R_{1,{\rm{noma}}}^{{x_1},\rm{m}} = \frac{1}{2}\ln \left( {1 + \gamma _{1,{\rm{noma}}}^{{x_1}}} \right), \,\,{\rm{when}} \,\,\, R_{1,{\rm{noma}}}^{{x_2},\rm{m}} \ge R_{{\rm{th}}}^{{x_2}},
	\label{snrU1x1ma}
\end{equation}
respectively, where
$\gamma _{1,{\rm{noma}}}^{{x_2}} = \frac{{{{a  _2}}{\rho _{\rm{s}}}{{\left\| {{\mathbf{h}_{{\rm{s}},1}}} \right\|}^2}}}{{{{{a  _1}}}{\rho _{\rm{s}}}{{\left\| {{\mathbf{h}_{{\rm{s}},1}}} \right\|}^2} + 1}}$,
$\gamma _{1,{\rm{noma}}}^{{x_1}} = {{{a  _1}}}{\rho _{\rm{s}}}{\left\| {{{\bf{h}}_{{\rm{s}},1}}} \right\|^2}$,
and the superscript `m' denotes multiple antennas scenarios.

Meanwhile, $x_2$ is directly decoded at $R$ and the achievable rate is expressed as
\begin{equation}
	R_{{\rm{r}},{\rm{noma}}}^{{x_2},\rm{m}} = \frac{1}{2}\ln \left( {1 + \gamma _{{\rm{r}},{\rm{noma}}}^{{x_2}}} \right),
	\label{snrRx2ma}
\end{equation}
where
$\gamma _{{\rm{r}},{\rm{noma}}}^{{x_2}} = \frac{{{{a  _2}}{\rho _{\rm{s}}}{Y_{{\rm{s}},{\rm{r}}}}}}{{{{{a  _1}}}{\rho _{\rm{s}}}{Y_{{\rm{s}},{\rm{r}}}} + 1}}$
and
${Y_{{\rm{s}},{\rm{r}}}} = \frac{{{{\left| {{{\bf{h}}_{{\rm{s}},{\rm{r}}}}{\bf{h}}_{{\rm{s}},1}^H} \right|}^2}}}{{{{\left\| {{{\bf{h}}_{{\rm{s}},1}}} \right\|}^2}}}$.

With the same method as (\ref{a1constraint2}), we obtain
\begin{equation}
	{a_1} \le \theta\left( {1 - \frac{{{\tau _1}}}{{{Y_{{\rm{s}},{\rm{r}}}}}}} \right),{Y_{{\rm{s}},{\rm{r}}}} > {\tau _1}.
	\label{a1constraint2ma}
\end{equation}

To enhance the performance of $U_2$ and minimize the interference from $R$ to $U_1$ in the second time slot, $R$ utilizes a directional antenna transmission and antenna gain $G$ is approximately denoted as \cite{LeiH2022IoT}
\begin{equation}
	G = \left\{ {\begin{array}{*{20}{c}}
			{{G_0,}}&{{\text{inside the mainlobe}},} \\
			{\eta {G_0},}&{{\text{outside the mainlobe}},}
	\end{array}} \right.
\end{equation}
where $G_0$ is antenna gain for the mainlobe and $\eta  < 1$ is the attenuating factor for the sidelobe gain.
The received signals at ${U_1}$ and $U_2$ in the second time slot are expressed as
\begin{equation}
	y_1^{{t_2}} = {{\bf{h}}_{{\rm{s}},1}}{\bf{w}}\sqrt {{{P_{\rm{s}}}}} {x_3} + {h_{{\rm{r}},1}}\sqrt {\eta G_0{P_{\rm{r}}}} {x_2} + n_1^{{t_2}},
	\label{yU1t2ma}
\end{equation}
\begin{equation}
	y_2^{{t_2}} = {h_{{\rm{r}},2}}\sqrt {G_0{P_{\rm{r}}}} {x_2} + n_2^{{t_2}},
	\label{yU2t2ma}
\end{equation}
respectively.

Since $x_2$ is decoded in the first slot at $U_1$ and can be deleted from $y_1^{{t_2}}$, then the achievable rate  of $x_3$ at $U_1$ is expressed as
\begin{equation}
	R_{1,{\rm{noma}}}^{{x_3},\rm{m}} = \frac{1}{2}\ln \left( {1 + \gamma_{1,{\rm{noma}}}^{{x_3}}} \right),
	\,\,{\rm{when}} \,\,\,R_{1,{\rm{noma}}}^{{x_2},\rm{m}} \ge R_{{\rm{th}}}^{{x_2}},
	\label{snrU1x3ma}
\end{equation}
where $\gamma_{1,{\rm{noma}}}^{{x_3}} = G_0{\rho _{\rm{s}}}{\left\| {{{\bf{h}}_{{\rm{s}},1}}} \right\|^2}$.

And the achievable rate  of $x_2$ at $U_2$ is expressed as
\begin{equation}
	R_{2,{\rm{noma}}}^{{x_2},\rm{m}} = \frac{1}{2}\ln \left( {1 + \gamma_{2,{\rm{noma}}}^{{x_2}}} \right),
	\,\,{\rm{when}} \,\,\, R_{{\rm{r,noma}}}^{{x_2},\rm{m}} \ge R_{{\rm{th}}}^{{x_2}},
	\label{snrU2x2ma}
\end{equation}
where
$\gamma_{2,{\rm{noma}}}^{{x_2}} = G_0{\rho _{\rm{r}}}{\left| {{h_{{\rm{r}},2}}} \right|^2}$
and
${\rho _{\rm{r}}} = \frac{{P_{\rm{r}}}}{\sigma ^2}$.

For the scenarios with ${Y_{{\rm{s}},{\rm{r}}}} < {\tau _{\rm{1}}} $, OMA scheme is ultilzed to transmit signals.
Specifically, the first time-slot ($t_1$) is divided into two equal parts and transmit $x_1$ and $x_2$ with MRT, respectively.
The corresponding achievable rate at $U_1$ and $R$ are expressed as
\begin{equation}
	R_{1,{\rm{oma}}}^{{x_1},\rm{m}} = \frac{1}{4}\ln \left( {1 + \gamma _{1,{\rm{oma}}}^{{x_1}}} \right),
	\label{snrU1x1OMAma}
\end{equation}
\begin{equation}
	R_{{\rm{r}},{\rm{oma}}}^{{x_2},\rm{m}} = \frac{1}{4}\ln \left( {1 + \gamma _{{\rm{r}},{\rm{oma}}}^{{x_2}}} \right),
	\label{snrRx2OMAma}
\end{equation}
respectively,
where
$\gamma _{1,\rm{oma}}^{{x_1}} = {\rho _{\rm{s}}}{\left\| {{{\bf{h}}_{{\rm{s}},1}}} \right\|^2}$,
$\gamma _{{\rm{r}},{\rm{oma}}}^{{x_2}} = {\rho _{\rm{s}}}{\left\| {{{\bf{h}}_{{\rm{s}},{\rm{r}}}}} \right\|^2}$,
and the pre-log factor of $\frac{1}{4}$ is multiplied since the first time slot is divided into two equal parts.

It must be noted that $x_2$ also can be received at $U_1$ in the scenarios with OMA scheme and the corresponding achievable rate at $U_1$ is expressed as
\begin{equation}
	R_{1,{\rm{oma}}}^{{x_2},{\rm{m}}} = \frac{1}{4}\ln \left( {1 + \gamma _{1,{\rm{oma}}}^{{x_2}}} \right),
	\label{snrU1x2OMAma}
\end{equation}
where
$\gamma _{1,{\rm{oma}}}^{{x_2}} = {\rho _{\rm{s}}}{Y_{{\rm{s}},1}}$
and
${Y_{{\rm{s}},{\rm{1}}}} = \frac{{{{\left| {{{\bf{h}}_{{\rm{s}},{\rm{1}}}}{\bf{h}}_{{\rm{s}},{\rm{r}}}^H} \right|}^2}}}{{{{\left\| {{{\bf{h}}_{{\rm{s}},{\rm{r}}}}} \right\|}^2}}}$.

In the second time slot, $R$ decodes and forwards $x_2$ to $U_2$ and $S$ transmits $x_3$ to $U_1$, the achievable rate at $U_1$ depends on whether $x_2$ can be decoded successfully at $U_1$, which is expressed as
\begin{equation}
	R_{1,{\rm{oma}}}^{{x_2},{\rm{m}}} > R_{{\rm{th}}}^{{x_2}} \Leftrightarrow {Y_{{\rm{s}},{\rm{1}}}} > {\tau _{\rm{2}}},
	\label{a1constraint3ma}
\end{equation}
where ${\tau _{\rm{2}}} = \frac{{\exp \left( {4R_{{\rm{th}}}^{{x_2}}} \right) - 1}}{{{\rho _{\rm{s}}}}}$.
Then the corresponding achievable rate at $U_1$ is expressed as
\begin{equation}
	R_{1,{\rm{oma}}}^{{x_3},{\rm{m}}} = \left\{ {\begin{array}{*{20}{c}}
			{\frac{1}{2}\ln \left( {1 + \gamma _{1,{\rm{oma}}}^{{x_3},{\rm{ms}}}} \right),}&{{Y_{{\rm{s}},{\rm{1}}}} > {\tau _{\rm{2}}},}\\
			{\frac{1}{2}\ln \left( {1 + \gamma _{1,{\rm{oma}}}^{{x_3},{\rm{mf}}}} \right),}&{{Y_{{\rm{s}},{\rm{1}}}} < {\tau _{\rm{2}}},}
	\end{array}} \right.
	\label{snrU1x3OMAma}
\end{equation}
where
$\gamma _{1,{\rm{oma}}}^{{x_3},{\rm{ms}}} = G_0{\rho _{\rm{s}}}{\left\| {{{\bf{h}}_{{\rm{s}},1}}} \right\|^2}$
and
$\gamma _{1,{\rm{oma}}}^{{x_3},{\rm{mf}}} = \frac{{{G_0\rho _{\rm{s}}}{{\left\| {{{\bf{h}}_{{\rm{s}},1}}} \right\|}^2}}}{{\eta G_0{\rho _{\rm{r}}}{{\left| {{h_{{\rm{r}},1}}} \right|}^2} + 1}}$,
the superscript `s' and `f' denotes success and failure, respectively.
The corresponding achievable rate at $U_2$ is expressed as
\begin{equation}
	R_{2,{\rm{oma}}}^{{x_2},\rm{m}} = \frac{1}{2}\ln \left( {1 + \gamma _{2,{\rm{oma}}}^{{x_2}}} \right),
	\label{snrU2x2OMAma}
\end{equation}
where $\gamma _{{2}, \rm {oma}}^{{x_2}} = G_0{\rho _{\rm{r}}}{\left| {{h_{{\rm{r}},2}}} \right|^2}$.

Based on \cite{VuT2022TVT} and \cite{LeiH2017TGCN}, the CDF and PDF of ${\left\| {{{\bf{h}}_{{\rm{s}},{\rm{d}}}}} \right\|^2}$  are expressed as
\begin{equation}
	{F_{{{\left\| {{{\bf{h}}_{{\rm{s}},{\rm{d}}}}} \right\|}^2}}}\left( y \right) = 1 - \exp \left( { - {\psi _{{\rm{s}},{\rm{d}}}}y} \right)\sum\limits_{m = 0}^{N - 1} {{\xi _{{\rm{s}},{\rm{d}}}}{y^m}},
	\label{cdf}
\end{equation}
\begin{equation}
	{f_{{{\left\| {{{\bf{h}}_{{\rm{s}},{\rm{d}}}}} \right\|}^2}}}\left( y \right) = \frac{{\psi _{{\rm{s}},{\rm{d}}}^N}}{{\Gamma \left( N \right)}}{y^{N - 1}}\exp \left( { - {\psi _{{\rm{s}},{\rm{d}}}}y} \right),
\end{equation}
where 
${\rm{d}} \in \left\{ {1,{\rm{r}}} \right\}$, 
${\psi _{{\rm{s}},{\rm{d}}}} = \frac{1}{{{\lambda _{{\rm{s}},{\rm{d}}}}}}$,
${\xi _{{\rm{s}},{\rm{d}}}} = \frac{{\psi _{{\rm{s}},{\rm{d}}}^m}}{m!}$,
and $\Gamma \left( x \right) $ is the gamma function, respectively.
The PDF of ${Y_{{\rm{s}},{\rm{r}}}}$ is given as \cite{VuT2022TVT}
\begin{equation}
	{f_{{Y_{{\rm{s}},{\rm{r}}}}}}\left( y \right) = {\psi _{{\rm{s}},{\rm{r}}}}\exp \left( { - {\psi _{{\rm{s}},{\rm{r}}}}y} \right).
	\label{pdf1}
\end{equation}

\section{Effective Sum Throughput Analysis}
\label{ESTAnalysis}

In this section, the scenario is considered wherein the traffic is the delay-sensitive and the signals are transmitted at the constant rate.
The effective sum throughput (EST) is utilized as the performance metric, which is expressed as \cite{ZhongC2015TCOM}
\begin{equation}
	\Psi  = \sum\limits_{i = 1}^3 {R_{{\rm{th}}}^{{x_i}}\left( {1 - P_{{\rm{out}}}^{{x_i}}} \right)}
	\label{TP}
\end{equation}
where $R_{{\rm{th}}}^{{x_i}}$ signifies the target threshold for $x_i$ and
${P_{{\rm{out}}}^{{x_i}}}$ is the OP of $x_i$, which is derived in the following subsections.

\subsection{Exact Outage Probability Analysis with the DPU scheme}

Substituting ${a_1} = \theta \left( {1 - \frac{{{\tau _{\rm{1}}}}}{{{Y_1}}}} \right)$ into $\gamma_1^{{x_1}}$, the OP of $x_1$ based on $R_1^{{x_2}} \ge R_{{\rm{th}}}^{{x_2}}$ in DPU scheme is obtained as
\begin{equation}
	\begin{aligned}
			P_{{\rm{out}}}^{{x_1},{\rm{DPU}}} &= 1 - \Pr \left\{ {{Y_1} > {\tau _1},R_1^{{x_2}} \ge R_{{\rm{th}}}^{{x_2}},R_1^{{x_1}} \ge R_{{\rm{th}}}^{{x_1}}} \right\}\\
			&= 1 - \Pr \left\{ {{Y_1} > {\tau _1},R_1^{{x_1}} \ge R_{{\rm{th}}}^{{x_1}}} \right\}\\
			&= 1 - \Pr \left\{ {{Y_1} \ge {A_0} + {\tau _{\rm{1}}}} \right\}\\
			&= 1 - \exp \left( { - {\psi _{{\rm{s}},1}}\left( {{A_0} + {\tau _{\rm{1}}}} \right)} \right),
		\label{OPx1DPU}
	\end{aligned}
\end{equation}
where ${A_0} = \frac{{{\theta _1}\left( {{\theta _2} + 1} \right)}}{{{\rho _{\rm{s}}}}}$ and
${\theta _1} = \exp \left( {2R_{{\rm{th}}}^{{x_1}}} \right) - 1$.
It must be noted that only when both two hops are not outage, $x_2$ can be successfully decoded at $U_2$. Thus, the OP of $x_2$ is expressed as
\begin{equation}
	\begin{aligned}
			&P_{{\rm{out}}}^{{x_2},{\rm{DPU}}} = 1 - \Pr \left\{ {{Y_1} > {\tau _1},R_{\rm{r}}^{{x_2}} \ge R_{{\rm{th}}}^{{x_2}},R_2^{{x_2}} \ge R_{{\rm{th}}}^{{x_2}}} \right\}\\
			&= 1 - \Pr \left\{ {{Y_1} > {\tau _1},\frac{{{\rho _{\rm{s}}}{a_2}{X_1}}}{{1 + {\rho _{\rm{s}}}{a_1}{X_1}}} \ge {\theta _2},{\rho _{\rm{r}}}{{\left| {{h_{{\rm{r}},2}}} \right|}^2} \ge {\theta _2}} \right\}\\
			&= 1 - \Pr \left\{ {{Y_1} > {\tau _1},{X_1} \ge {Y_1},{{\left| {{h_{{\rm{r}},2}}} \right|}^2} \ge \frac{{{\theta _2}}}{{{\rho _{\rm{r}}}}}} \right\}\\
			&= 1 - \exp \left( { - \frac{{{\psi _{{\rm{r,2}}}}{\theta _2}}}{{{\rho _{\rm{r}}}}}} \right)\int_{{\tau _1}}^\infty  {\left( {1 - {F_{{X_1}}}\left( y \right)} \right){f_{{Y_1}}}\left( y \right)dy} \\
			&= 1 - \frac{{{\psi _{{\rm{s,1}}}}}}{{{\psi _{{\rm{s,r}}}} + {\psi _{{\rm{s,1}}}}}}\exp \left( { - \left( {{\psi _{{\rm{s}},{\rm{r}}}} + {\psi _{{\rm{s}},{\rm{1}}}}} \right){\tau _1} - \frac{{{\psi _{{\rm{r}},{\rm{2}}}}{\theta _2}}}{{{\rho _{\rm{r}}}}}} \right),
		\label{OPx2DPU}
	\end{aligned}
\end{equation}
Similarly, it is worth noting that $R_1^{{x_2}} \ge R_{{\rm{th}}}^{{x_2}}$ is satisfied, $x_3$ can be decoded without interference. Thus, the OP of $x_3$ is obtained as
\begin{equation}
	\begin{aligned}
			P_{{\rm{out}}}^{{x_3},{\rm{DPU}}} &= 1 - \Pr \left\{ {{Y_1} > {\tau _1},R_1^{{x_2}} \ge R_{{\rm{th}}}^{{x_2}},R_1^{{x_3}} \ge R_{{\rm{th}}}^{{x_3}}} \right\}\\
			&= 1 - \Pr \left\{ {{Y_1} > {\tau _1},R_1^{{x_3}} \ge R_{{\rm{th}}}^{{x_3}}} \right\}\\
			&= 1 - \Pr \left\{ {{Y_1} > \max \left( {{\tau _1},\frac{{{\theta _3}}}{{{\rho _{\rm{s}}}}}} \right)} \right\}\\
			&= \left\{ {\begin{array}{*{20}{c}}
					{1 - \exp \left( { - {\psi _{{\rm{s}},1}}{\tau _1}} \right),}&{{\theta _2} > {\theta _3}},\\
					{1 - \exp \left( { - \frac{{{\psi _{{\rm{s}},1}}{\theta _3}}}{{{\rho _{\rm{s}}}}}} \right),}&{{\theta _2} < {\theta _3}},
			\end{array}} \right.
		\label{OPx2DPU}
	\end{aligned}
\end{equation}
where
${\theta _3} = \exp \left( {2R_{{\rm{th}}}^{{x_3}}} \right) - 1$.

\subsection{Exact Outage Probability Analysis with DPR scheme}

With the same method as (\ref{OPx1DPU}) and by utilizing
\begin{equation}
	\int_{{x_0}}^{{x_1}} {e^ {{ - ax - \frac{b}{x}}}dx}  \approx \left( {{e^{ - a{x_0}}} - {e^{ - a{x_1}}}} \right)\sqrt {\frac{{4b}}{a}} {K_1}\left( {\sqrt {4ab} } \right),
	\label{YanM2012WCL}
\end{equation}
which is verified in \cite{YanM2012WCL}, the OP of $x_1$ in DPR scheme is obtained as
\begin{equation}
	\begin{aligned}
			P_{{\rm{out}}}^{{x_1},{\rm{DPR}}} &= 1 - \Pr \left\{ {{X_1} > {\tau _1},R_1^{{x_2}} \ge R_{{\rm{th}}}^{{x_2}},R_1^{{x_1}} \ge R_{{\rm{th}}}^{{x_1}}} \right\}\\
			&= 1 - \Pr \left\{ {{X_1} > {\tau _1},\frac{{{\rho _{\rm{s}}}{a_2}{Y_1}}}{{1 + {\rho _{\rm{s}}}{a_1}{Y_1}}} \ge {\theta _2},{\rho _{\rm{s}}}{a_1}{Y_1} \ge {\theta _1}} \right\}\\
			&= 1 - \Pr \left\{ {{X_1} > {\tau _1},{Y_1} \ge \max \left( {{X_1},\frac{{{A_0}{X_1}}}{{{X_1} - {\tau _1}}}} \right)} \right\}\\
			&= 1 - \Pr \left\{ {{Y_1} \ge {X_1},{X_1} > {A_0} + {\tau _1}} \right\} \\
			&\;\;\;\;\;\;- \Pr \left\{ {{Y_1} \ge \frac{{{A_0}{X_1}}}{{{X_1} - {\tau _1}}},{\tau _1} < {X_1} < {A_0} + {\tau _1}} \right\}\\
			&= 1 - \frac{{{\psi _{{\rm{s,r}}}}}}{{{\psi _{{\rm{s,1}}}} + {\psi _{{\rm{s,r}}}}}}\exp \left( { - \left( {{\psi _{{\rm{s,1}}}} + {\psi _{{\rm{s,r}}}}} \right)\left( {{A_0} + {\tau _1}} \right)} \right)\\
			&\;\;\;\;\;\; - {\psi _{{\rm{s,r}}}}\exp \left( { - {\psi _{{\rm{s,r}}}}{\tau _1} - {\psi _{{\rm{s,1}}}}{A_0}} \right)\\
			&\;\;\;\;\;\; \times \int_0^{{A_0}} {\exp \left( { - \frac{{{\psi _{{\rm{s,1}}}}{A_0}{\tau _1}}}{t}} \right)\exp \left( { - {\psi _{{\rm{s,r}}}}t} \right)dt} \\
			&\approx 1 - \frac{{{\psi _{{\rm{s,r}}}}}}{{{\psi _{{\rm{s,1}}}} + {\psi _{{\rm{s,r}}}}}}\exp \left( { - \left( {{\psi _{{\rm{s,1}}}} + {\psi _{{\rm{s,r}}}}} \right)\left( {{A_0} + {\tau _1}} \right)} \right)\\
			&\;\;\;\;\;\;- \exp \left( { - {\psi _{{\rm{s,r}}}}{\tau _1} - {\psi _{{\rm{s,1}}}}{A_0}} \right)\left( {1 - \exp \left( { - {\psi _{{\rm{s,r}}}}{A_0}} \right)} \right)\\
			&\;\;\;\;\;\; \times {\phi _1}\left( {4{\psi _{{\rm{s,1}}}}{\psi _{{\rm{s,r}}}}{\tau _1}{A_0}} \right),
		\label{OPx1DPR}
	\end{aligned}
\end{equation}
where ${\phi _1}\left( x \right) = \sqrt x {K_1}\left( {\sqrt x } \right)$ and ${{K_v}\left(  \cdot  \right)}$ is the $vth$-order modified Bessel function of the second kind, defined by \cite[(8.432.1)]{Gradshteyn2007Book}.

Similar as (\ref{OPx2DPU}), it is worth noting that the implementation of the DPA scheme at $R$ ensures that $R_{\rm{r}}^{{x_2}} \ge R_{{\rm{th}}}^{{x_2}}$ is satisfied. Thus, the OP of $x_2$ in DPR scheme is expressed as
\begin{equation}
	\begin{aligned}
	P_{{\rm{out}}}^{{x_2},{\rm{DPR}}} &= 1 - \Pr \left\{ {{X_1} > {\tau _1},R_{\rm{r}}^{{x_2}} \ge R_{{\rm{th}}}^{{x_2}},R_2^{{x_2}} \ge R_{{\rm{th}}}^{{x_2}}} \right\}\\
	&= 1 - \Pr \left\{ {{X_1} > {\tau _1},{{\left| {{h_{{\rm{r}},2}}} \right|}^2} \ge \frac{{{\theta _2}}}{{{\rho _{\rm{r}}}}}} \right\}\\
	&= 1 - \exp \left( { - {\psi _{{\rm{s}},{\rm{r}}}}{\tau _1} - \frac{{{\psi _{{\rm{r}},{\rm{2}}}}{\theta _2}}}{{{\rho _{\rm{r}}}}}} \right).
		\label{OPx2DPR}
\end{aligned}
\end{equation}

The OP of $x_3$ in DPR scheme is expressed as (\ref{OPx3DPR}), shown at the top of the next page.

\begin{figure*}[ht]
\begin{equation}
	\begin{aligned}
			P_{{\rm{out}}}^{{x_3},{\rm{DPR}}} &= 1 - \Pr \left\{ {{X_1} > {\tau _1},R_1^{{x_2}} \ge R_{{\rm{th}}}^{{x_2}},R_1^{{x_3}} \ge R_{{\rm{th}}}^{{x_3}}} \right\}\\
			&= 1 - \Pr \left\{ {{X_1} > {\tau _1},{Y_1} \ge \max \left( {{X_1},\frac{{{\theta _3}}}{{{\rho _{\rm{s}}}}}} \right)} \right\}\\
			&= 1 - \Pr \left\{ {{X_1} > {\tau _1},{Y_1} \ge {X_1},{X_1} > \frac{{{\theta _3}}}{{{\rho _{\rm{s}}}}}} \right\} - \Pr \left\{ {{X_1} > {\tau _1},{Y_1} \ge \frac{{{\theta _3}}}{{{\rho _{\rm{s}}}}},{X_1} < \frac{{{\theta _3}}}{{{\rho _{\rm{s}}}}}} \right\}\\
			&= \left\{ {\begin{array}{*{20}{c}}
					{1 - \frac{{{\psi _{{\rm{s,r}}}}}}{{{\psi _{{\rm{s,1}}}} + {\psi _{{\rm{s,r}}}}}}\exp \left( { - \left( {{\psi _{{\rm{s,1}}}} + {\psi _{{\rm{s,r}}}}} \right){\tau _1}} \right),}&{{\theta _2} > {\theta _3}}\\
					{1 - \exp \left( { - {\tau _1}{\psi _{{\rm{s,r}}}} - \frac{{{\theta _3}{\psi _{{\rm{s,1}}}}}}{{{\rho _{\rm{s}}}}}} \right) + \frac{{{\psi _{{\rm{s,1}}}}}}{{{\psi _{{\rm{s,1}}}} + {\psi _{{\rm{s,r}}}}}}\exp \left( { - \frac{{{\theta _3}\left( {{\psi _{{\rm{s,1}}}} + {\psi _{{\rm{s,r}}}}} \right)}}{{{\rho _{\rm{s}}}}}} \right),}&{{\theta _2} < {\theta _3}}
			\end{array}} \right.
		\label{OPx3DPR}
	\end{aligned}
\end{equation}
\hrulefill
\end{figure*}

\subsection{Exact Outage Probability Analysis with MDPR scheme}

Substituting ${{a  _1}} = \theta \left( {1 - \frac{{\tau _{\rm{1}}} }{{{Y_{{\rm{s}},{\rm{r}}}}}}} \right)$ into $\gamma_{1,{\rm{noma}}}^{{x_1}}$ and utilizing \cite[(3.471.9)]{Gradshteyn2007Book}, the OP of $x_1$ in MDPR scheme is obtained as (\ref{OPx1MDPR}), shown at the top of this page, 
where ${A_1} = \left( {1 - \exp \left( { - {\psi _{{\rm{s}},{\rm{r}}}}{\tau _{\rm{1}}} } \right)} \right)\xi _{{\rm{s}},1}{\rho _{\rm{s}}}^{ - {m}}$, ${A_2} = {\rm{ }}\frac{{{\xi _{{\rm{s}},1}}\left( {_n^m} \right)}}{{{{\left( {\theta {\rho _{\rm{s}}}} \right)}^m}}}{\left( {\frac{{{\psi _{{\rm{s}},{\rm{r}}}}\theta {\theta _2}}}{{{\psi _{{\rm{s}},1}}}}} \right)^{\frac{n}{2}}}\exp \left( { - {\psi _{{\rm{s}},{\rm{r}}}}{\tau _1}} \right)$, 
and 
${A_3} = \frac{{{\psi _{{\rm{s}},1}}}}{{\theta {\rho _{\rm{s}}}}}$.

\begin{figure*}[ht]
\begin{equation}
	\begin{aligned}
			P_{{\rm{out}}}^{{x_1},{\rm{MDPR}}} &= \Pr \left\{ {{Y_{{\rm{s}},{\rm{r}}}} > {\tau _1},R_{1,{\rm{noma}}}^{{x_2},{\rm{m}}} > R_{{\rm{th}}}^{{x_2}},R_{1,{\rm{noma}}}^{{x_1},{\rm{m}}} < R_{{\rm{th}}}^{{x_1}}} \right\} \\
			&\;+ \Pr \left\{ {{Y_{{\rm{s}},{\rm{r}}}} < {\tau _1},R_{1,{\rm{oma}}}^{{x_1},{\rm{m}}} < R_{{\rm{th}}}^{{x_1}}} \right\}\\
			&= \Pr \left\{ {{Y_{{\rm{s}},{\rm{r}}}} > {\tau _1},R_{1,{\rm{noma}}}^{{x_1},{\rm{m}}} < R_{{\rm{th}}}^{{x_1}}} \right\} + \Pr \left\{ {{Y_{{\rm{s}},{\rm{r}}}} < {\tau _1},R_{1,{\rm{oma}}}^{{x_1},{\rm{m}}} < R_{{\rm{th}}}^{{x_1}}} \right\}\\
			&= \Pr \left\{ {{Y_{{\rm{s}},{\rm{r}}}} > {\tau _1},\gamma _{1,{\rm{noma}}}^{{x_1}} < {\theta _1}} \right\} + \Pr \left\{ {{Y_{{\rm{s}},{\rm{r}}}} < {\tau _1},\gamma _{1,{\rm{oma}}}^{{x_1}} < {\theta _1}\left( {{\theta _1} + 2} \right)} \right\}\\
			&= \Pr \left\{ {{Y_{{\rm{s}},{\rm{r}}}} > {\tau _1},{{\left\| {{{\bf{h}}_{{\rm{s}},{\rm{1}}}}} \right\|}^2} < \frac{{{\theta _1}}}{{\theta {\rho _{\rm{s}}}}}\frac{{{Y_{{\rm{s}},{\rm{r}}}}}}{{{Y_{{\rm{s}},{\rm{r}}}} - {\tau _{\rm{r}}}}}} \right\} \\
			&\;\;\;\;+ {F_{{Y_{{\rm{s}},{\rm{r}}}}}}\left( {{\tau _1}} \right)\Pr \left\{ {{{\left\| {{{\bf{h}}_{{\rm{s}},{\rm{1}}}}} \right\|}^2} < \frac{{{\theta _1}\left( {{\theta _1} + 2} \right)}}{{{\rho _{\rm{s}}}}}} \right\}\\
			&= \int_{{\tau _{\rm{r}}}}^\infty  {{F_{{{\left\| {{{\bf{h}}_{{\rm{s}},{\rm{1}}}}} \right\|}^2}}}\left( {\frac{{{\theta _1}}}{{\theta {\rho _{\rm{s}}}}}\frac{y}{{y - {\tau _1}}}} \right){f_{{Y_{{\rm{s}},{\rm{r}}}}}}\left( y \right)dy}  + {F_{{Y_{{\rm{s}},{\rm{r}}}}}}\left( {{\tau _1}} \right){F_{{{\left\| {{{\bf{h}}_{{\rm{s}},{\rm{1}}}}} \right\|}^2}}}\left( {\frac{{{\theta _1}\left( {{\theta _1} + 2} \right)}}{{{\rho _{\rm{s}}}}}} \right)\\
			&= 1 - \sum\limits_{m = 0}^{N - 1} {{A_1}{{\left( {{\theta _1}\left( {{\theta _1} + 2} \right)} \right)}^m}\exp \left( { - {A_3}\theta {\theta _1}\left( {{\theta _1} + 2} \right)} \right)} - \sum\limits_{m = 0}^{N - 1} {\sum\limits_{n = 0}^m {{A_2}\theta _1^{m - \frac{n}{2}}\exp \left( { - {A_3}{\theta _1}} \right){\phi _1}\left( {4{A_3}{\psi _{{\rm{s}},{\rm{r}}}}{\theta _1}{\tau _1}} \right)} }
		\label{OPx1MDPR}
	\end{aligned}
\end{equation}
	\hrulefill
\end{figure*}


It must be noted that only when both two hops are not outage, $x_2$ can be successfully decoded at $U_2$. Thus, the OP of $x_2$ in MDPR scheme is obtained as 
(\ref{OPx2}), shown at the top of this page, 
where ${A_{4}} = \left( {1 - \exp \left( { - {\psi _{\rm{s,r}}}{\tau _{\rm{1}}} } \right)} \right)\xi _{\rm{s,r}}{{\rho _{\rm{s}}}}^{ - m}$,
${A_5} = \frac{{{\psi _{\rm{s,r}}}}}{{{{\rho _{\rm{s}}}}}}$, and ${A_6} = \frac{{{\psi _{{\rm{r}},{\rm{2}}}}}}{{{G_0}{\rho _{\rm{r}}}}}$.

\begin{figure*}[ht]
\begin{equation}
	\begin{aligned}
			P_{{\rm{out}}}^{{x_2},{\rm{MDPR}}} &= 1 - \Pr \left\{ {{Y_{{\rm{s}},{\rm{r}}}} > {\tau _1},R_{2,{\rm{noma}}}^{{x_2},{\rm{m}}} > R_{{\rm{th}}}^{{x_2}},R_{{\rm{r}},{\rm{noma}}}^{{x_2},{\rm{m}}} > R_{{\rm{th}}}^{{x_2}}} \right\}\\
			&\;\;\;\;\;\; - \Pr \left\{ {{Y_{{\rm{s}},{\rm{r}}}} < {\tau _1},R_{2,{\rm{oma}}}^{{x_2},{\rm{m}}} > R_{{\rm{th}}}^{{x_2}},R_{{\rm{r}},{\rm{oma}}}^{{x_2},{\rm{m}}} > R_{{\rm{th}}}^{{x_2}}} \right\}\\
			&= 1 - \Pr \left\{ {{Y_{{\rm{s}},{\rm{r}}}} > {\tau _1},R_{{\rm{2}},{\rm{noma}}}^{{x_2},{\rm{m}}} > R_{{\rm{th}}}^{{x_2}}} \right\}\\
			&\;\;\;\;\;\;  - \Pr \left\{ {{Y_{{\rm{s}},{\rm{r}}}} < {\tau _1},R_{2,{\rm{oma}}}^{{x_2},{\rm{m}}} > R_{{\rm{th}}}^{{x_2}},R_{{\rm{r}},{\rm{oma}}}^{{x_2},{\rm{m}}} > R_{{\rm{th}}}^{{x_2}}} \right\}\\
			&= 1 - \exp \left( { - {\psi _{{\rm{s}},{\rm{r}}}}{\tau _1}} \right)\exp \left( { - {A_6}{\theta _2}} \right) - \sum\limits_{m = 0}^{N - 1} {{A_4}{{\left( {{\theta _2}\left( {{\theta _2} + 2} \right)} \right)}^m}\exp \left( { - \left( {\left( {{\theta _2} + 2} \right){A_5} + {A_6}} \right){\theta _2}} \right)}
		\label{OPx2}
	\end{aligned}
\end{equation}
	\hrulefill
\end{figure*}

Similarly, utilizing \cite[(3.351.3)]{Gradshteyn2007Book}, the OP of $x_3$ in MDPR scheme is obtained as 
(\ref{OPx3}), shown at the top of the next page, 
where ${A_7} = \xi _{{\rm{s}},1}\exp \left( { - {\psi _{{\rm{s}},{\rm{r}}}}{\tau _{\rm{1}}} } \right){\rho _{\rm{s}}}^{ - {m}}$, ${A_8} = \left( {1 - \exp \left( { - {\psi _{{\rm{s}},{\rm{r}}}}{\tau _{\rm{1}}}} \right)} \right){\xi _{{\rm{s}},1}}{\rho _{\rm{s}}}^{ - m}{\psi _{{\rm{r}},1}}n!{\left( {\eta {G_0}{\rho _{\rm{r}}}} \right)^n}\left( {_n^m} \right){\left( {\frac{{{A_9}}}{{{\psi _{{\rm{r}},1}}}}} \right)^{n + 1}}$, ${A_9} = \frac{{{\psi _{{\rm{r}},1}}{\rho _{\rm{s}}}}}{{{\psi _{{\rm{s}},1}}\eta {\rho _{\rm{r}}}}}$, and ${A_{10}} = 1 - \exp \left( { - {\psi _{{\rm{s}},1}}{\tau _2}} \right)$.

\begin{figure*}[ht]
\begin{equation}
	\begin{small}
	\begin{aligned}
		P_{{\rm{out}}}^{{x_3},{\rm{MDPR}}} &= \Pr \left\{ {{Y_{{\rm{s}},{\rm{r}}}} > {\tau _1},R_{1,{\rm{noma}}}^{{x_2},{\rm{m}}} > R_{{\rm{th}}}^{{x_2}},R_{1,{\rm{noma}}}^{{x_3},{\rm{m}}} < R_{{\rm{th}}}^{{x_3}}} \right\} \\
		&\;+ \Pr \left\{ {{Y_{{\rm{s}},{\rm{r}}}} < {\tau _1},R_{1,{\rm{oma}}}^{{x_3},{\rm{m}}} < R_{{\rm{th}}}^{{x_3}}} \right\}\\
		&= \Pr \left\{ {{Y_{{\rm{s}},{\rm{r}}}} > {\tau _1},R_{1,{\rm{noma}}}^{{x_3},{\rm{m}}} < R_{{\rm{th}}}^{{x_3}}} \right\} + \Pr \left\{ {{Y_{{\rm{s}},{\rm{r}}}} < {\tau _1},R_{1,{\rm{oma}}}^{{x_3},{\rm{m}}} < R_{{\rm{th}}}^{{x_3}}} \right\}\\
		&= \Pr \left\{ {{Y_{{\rm{s}},{\rm{r}}}} > {\tau _1},\gamma _{{\rm{1}},{\rm{noma}}}^{{x_3}} < {\theta _3}} \right\} + \Pr \left\{ {{Y_{{\rm{s}},{\rm{r}}}} < {\tau _1},{Y_{{\rm{s}},{\rm{1}}}} > {\tau _2},\gamma _{1,{\rm{oma}}}^{{x_3},{\rm{ms}}} < {\theta _3}} \right\}\\
		&\;+ \Pr \left\{ {{Y_{{\rm{s}},{\rm{r}}}} < {\tau _1},{Y_{{\rm{s}},{\rm{1}}}} < {\tau _2},\gamma _{1,{\rm{oma}}}^{{x_3},{\rm{mf}}} < {\theta _3}} \right\}\\
		&= \Pr \left\{ {{Y_{{\rm{s}},{\rm{r}}}} > {\tau _1},{{\left\| {{{\bf{h}}_{{\rm{s}},{\rm{1}}}}} \right\|}^2} < \frac{{{\theta _3}}}{{{G_0}{\rho _{\rm{s}}}}}} \right\} + \Pr \left\{ {{Y_{{\rm{s}},{\rm{r}}}} < {\tau _1},{Y_{{\rm{s}},{\rm{1}}}} > {\tau _2},{{\left\| {{{\bf{h}}_{{\rm{s}},{\rm{1}}}}} \right\|}^2} < \frac{{{\theta _3}}}{{{G_0}{\rho _{\rm{s}}}}}} \right\}\\
		&\;+ \Pr \left\{ {{Y_{{\rm{s}},{\rm{r}}}} < {\tau _1},{Y_{{\rm{s}},{\rm{1}}}} < {\tau _2},{{\left\| {{{\bf{h}}_{{\rm{s}},{\rm{1}}}}} \right\|}^2} < \frac{{\left( {\eta {G_0}{\rho _{\rm{r}}}{{\left| {{h_{{\rm{r}},{\rm{1}}}}} \right|}^2} + 1} \right){\theta _3}}}{{{G_0}{\rho _{\rm{s}}}}}} \right\}\\
		&= 1 - \exp \left( { - {\psi _{{\rm{s}},1}}{\tau _2}} \right)\sum\limits_{m = 0}^{N - 1} {{\xi _{{\rm{s}},1}}{{\left( {\frac{{{\theta _3}}}{{{G_0}{\rho _{\rm{s}}}}}} \right)}^m}} \exp \left( { - \frac{{{\psi _{{\rm{s}},1}}{\theta _3}}}{{{G_0}{\rho _{\rm{s}}}}}} \right)\\
		&- \sum\limits_{m = 0}^{N - 1} {{A_{10}}{{\left( {\frac{{{\theta _3}}}{{{G_0}}}} \right)}^m}\exp \left( { - \frac{{{\psi _{{\rm{s}},1}}{\theta _3}}}{{{G_0}{\rho _{\rm{s}}}}}} \right)} \left( {{A_7} + \sum\limits_{n = 0}^m {{A_8}} {{\left( {{\theta _3} + {A_9}} \right)}^{ - n - 1}}} \right)
		\label{OPx3}
	\end{aligned}
   \end{small}
\end{equation}
	\hrulefill
\end{figure*}

\begin{table*}[!t]
	\caption{Comparisons of Schemes.}
	\center
	\begin{tabular}{|c | c | c| c |}
		\hline
		$\mathrm{Scheme}$  & Antenna at $S$ & Power Allocation Scheme & Beamforming scheme \\
		
		\hline
		DPU     & Single  & DPA based on $S$-$U_1$    & ---\\
		\hline
		DPR     & Single  & DPA based on $S$-$R$      & ---\\
		\hline			
		Ben1    & Single  & FPA               & ---\\
		\hline	
		Ben2    & Multiple  & FPA               & based on $S$-$U_1$\\
		\hline
		Ben3     & Multiple & DPA based on $S$-$U_1$    & based on $S$-$R$\\
		\hline
		The proposed (MDPR)    & Multiple & DPA  based on $S$-$R$      & based on $S$-$U_1$\\
		\hline
	\end{tabular}
	\label{tableI}
\end{table*}

\section{Simulation Results}
\label{sec: RESULTS}

Simulation results are presented in this section to validate the proposed scheme's effectiveness.
The effects of system parameters on the performance of the considered scheme, such as normalized power, the distance between the transmitter and receiver, and power allocation coefficients, are investigated.
The main parameters are set as follows:
${d_{{\rm{s,1}}}} = 10$ m, ${d_{{\rm{s,r}}}} = 15 = {d_{{\rm{r,1}}}} = 15$ m, ${d_{{\rm{r,2}}}} = 10$ m, $\alpha = 2$, $R_{{\rm{th}}}^{{x_1}} = R_{{\rm{th}}}^{{x_2}} =  R_{{\rm{th}}}^{{x_3}} = R_{{\rm{th}}} = 0.2$ nat/s/Hz, $\eta = 0.7$, and $N = 10$, respectively.
In all the figures, `Sim' and `Ana' denote the simulation and numerical, respectively.
One can observe that the simulation and numerical results match perfectly to verify the correctness of the analysis.

The following three schemes are utilized as benchmarks to prove the superiority of the proposed schemes:
\begin{enumerate}
	\item A NOMA-based CDRT system with FPA scheme (`Ben1') \cite{KimJ2015CL}:
	In this scheme, the condition of whether $R$ can decode the information of $U_2$ was not considered and all the nodes were equipped with a single antenna.
	
	\item A NOMA-based CDRT system with FPA and beamforming schemes (`Ben2'):
	In this scheme, the FPA scheme is utilized at $S$ and the beamforming scheme is designed based on $S$-$R$ link.
	
	\item A NOMA-based CDRT system with DPA and beamforming schemes (`Ben3'):
	In this scheme, the beamforming scheme is designed based on $S$-$R$ link and the DPA scheme is based on $S$-$U_1$.
\end{enumerate}
Table \ref{tableI} summarizes the difference for all the schemes.

\begin{figure}[!t]
	\centering
	\subfigure[OP of the $x_1$]{
		\label{fig21}
		\includegraphics[width = 0.3 \textwidth]{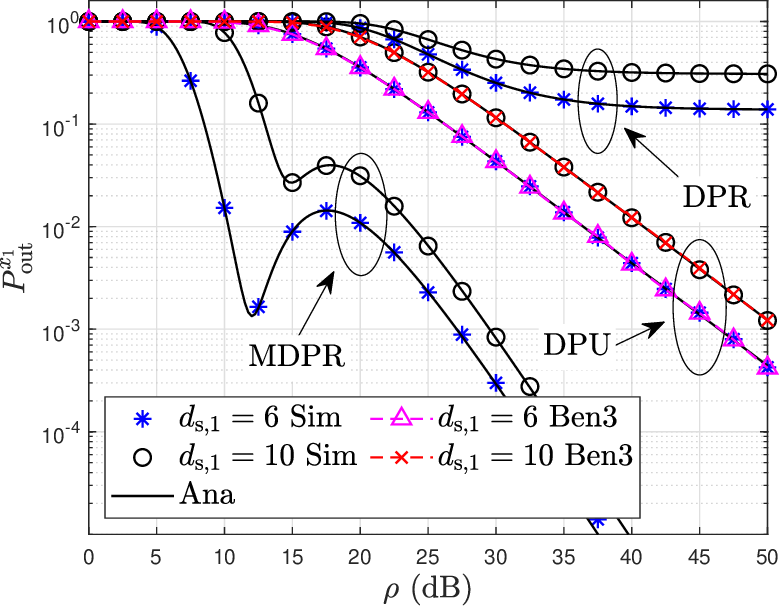}}
	\subfigure[OP of $x_2$]{
		\label{fig22}
		\includegraphics[width = 0.3 \textwidth]{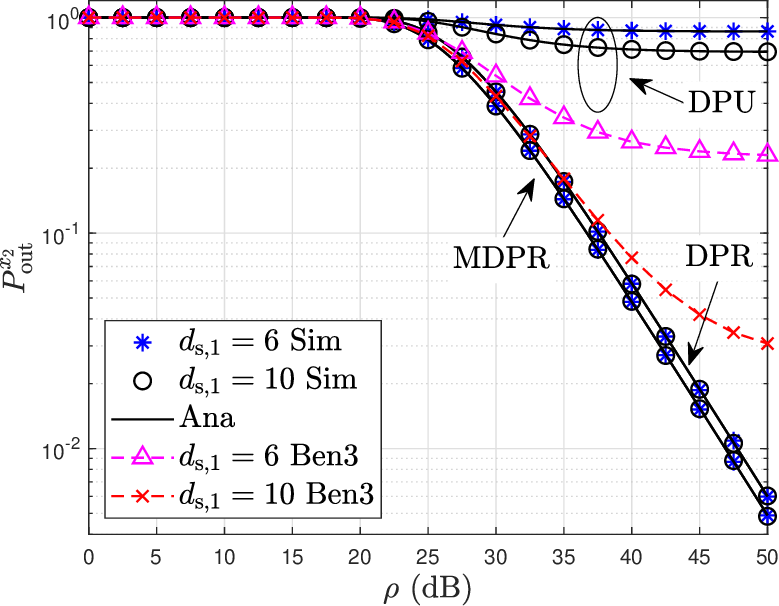}}
	\subfigure[OP of $x_3$]{
		\label{fig23}
		\includegraphics[width = 0.3 \textwidth]{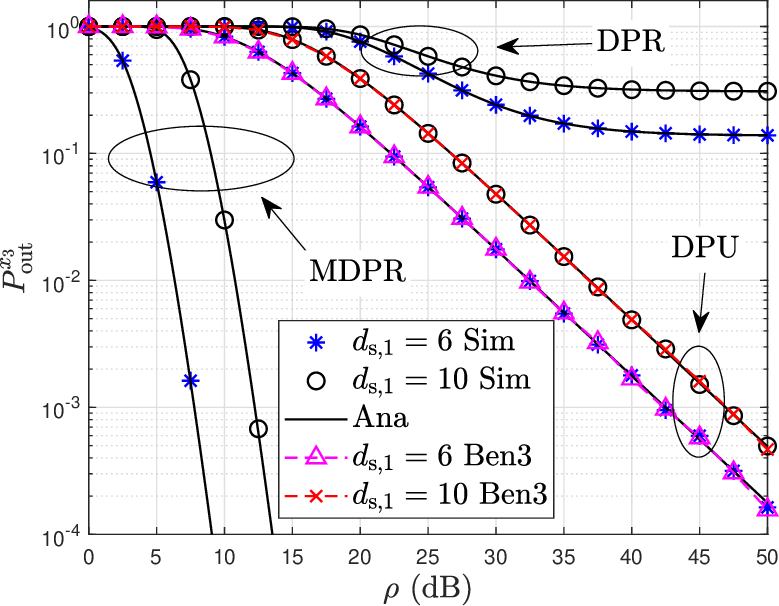}}
	\subfigure[EST]{
		\label{fig24}
		\includegraphics[width = 0.3 \textwidth]{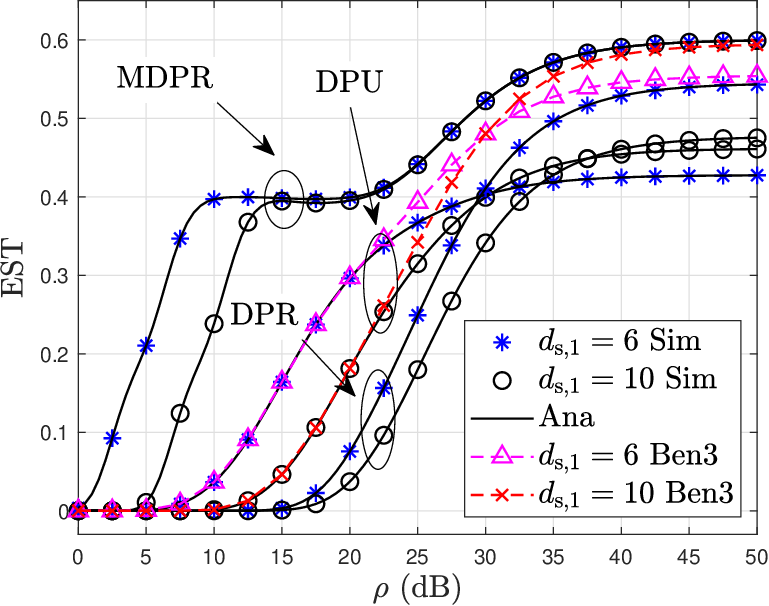}}
	\caption{OPs and EST for varying ${\rho}$ and ${d_{{\rm{s}},1}}$.}
	\label{fig2}
\end{figure}
Fig. \ref{fig2} presents the impact of ${\rho}$ and varying ${d_{{\rm{s}},1}}$ on the OPs and EST.
In Fig.  \ref{fig21}, the OP of $x_1$ in the DPR scheme decreases and remains constant with $\rho$ increasing.
This is because the first step of the SIC on $U_1$ is not guaranteed; therefore, the OP of $x_1$ is dependent on whether $R_1 ^{x_2}$ is an outage or not.
Unlike the DPR scheme, the OP of $x_1$ decreases in the DPU scheme with $\rho$ increases because the first step of the SIC on $U_1$ is invariably satisfied.
The OP of DPU and Ben3 is almost equal; this is because the beamforming scheme in Ben3 is based on $S$-$R$ link and the power allocation is based on $ S$-$U_1$ link.
In the MDPR scheme, the OP of $x_1$ initially decreases, subsequently increases, and decreases with $\rho$ increases.
The reason is that the OMA scheme dominates in the lower-${\rho}$; as ${\rho}$ increases, NOMA is utilized.
One can find that the MDPR scheme has the best outage performance for $x_1$, which is since the beamforming scheme is based on $S$-$U_1$ and the power allocation is based on $ S$-$R$ ensures that $x_2$ can be decoded on $R$.
In Fig.  \ref{fig22}, we can observe that $P_{{\rm{out}}}^{{x_2}}$ with the DPU scheme decreases with $\rho$ until it becomes a constant.
Unlike \ref{fig21}, the outage performance for the DPR scheme outperforms that for the DPU scheme because that $R$ can successfully decode $x_2$ in the DPR scheme.
Similar to \ref{fig21}, the MDPR scheme performs best because the decoding at $R$ and $U_1$ are considered.
The outage performance of $x_2$ for Ben3 is also reduced until a constant, which outperforms that of the DPU but underreports that of the DPR. 
This is because the DPA in Ben3 is based on $S$-$U_1$ and the beamforming scheme is utilized on $S$-$R$, respectively.
Comparing the MDPR with Ben3, one can observe that DPA is more effective in enhancing the outage performance.
Similar to Fig. \ref{fig21}, the OP of $x_3$ for the DPU and MDPR schemes decreases with $\rho$ increases in Fig. \ref{fig23}.
However, $P_{{\rm{out}}}^{{x_3}}$ of the MDPR scheme decreases rapidly with increasing $\rho$ because $x_3$ is not affected by $x_2$.
Fig.  \ref{fig24} demonstrates the superior performance of the MDPR scheme against others due to the drastically improved outage performance of $x_1$ and $x_3$.
In Fig.  \ref{fig2}, we can also observe that, 
as $d_{\rm{s,1}}$ increases, the path loss between $S$ and $U_1$ increases and the outage performance of $x_1$ and $x_3$ become worse.

\begin{figure}[!t]
	\centering
	\subfigure[OP of $x_1$]{
		\label{fig31}
		\includegraphics[width = 0.3 \textwidth]{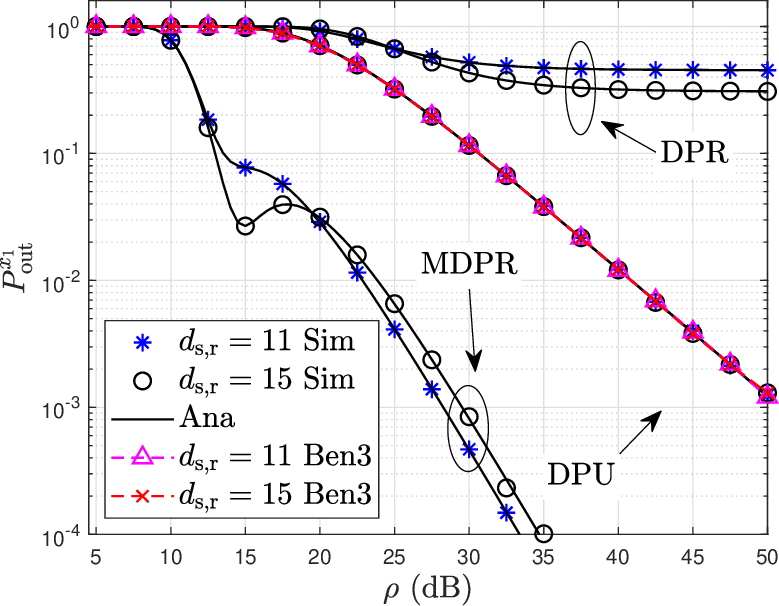}}
	\subfigure[OP of $x_2$]{
		\label{fig32}
		\includegraphics[width = 0.3 \textwidth]{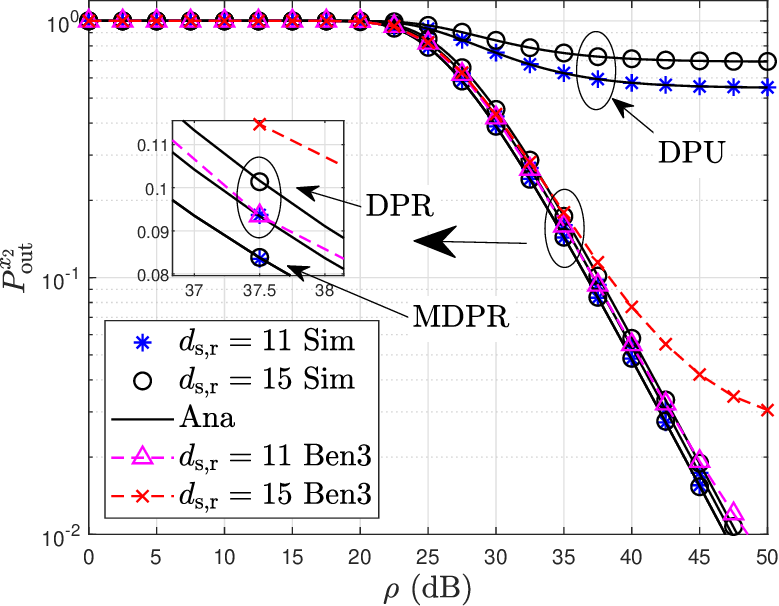}}
	\subfigure[OP of $x_3$]{
		\label{fig33}
		\includegraphics[width = 0.3 \textwidth]{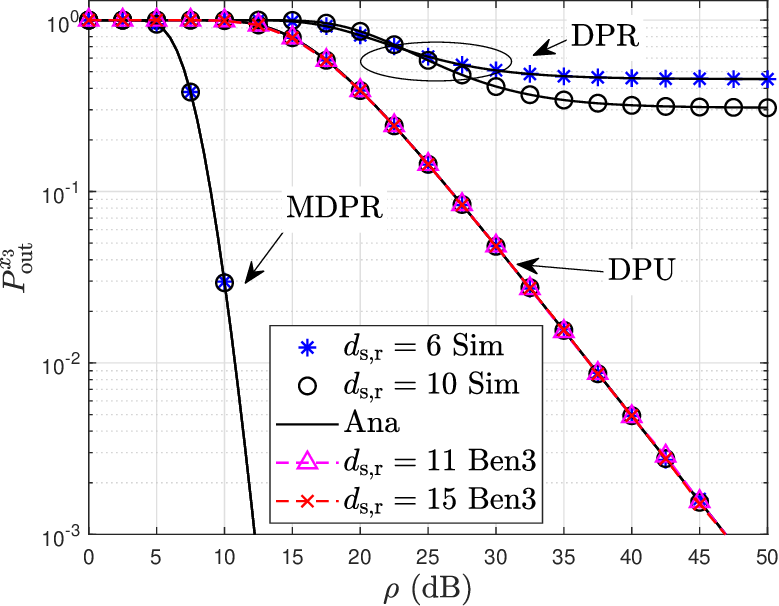}}
	\subfigure[EST]{
		\label{fig34}
		\includegraphics[width = 0.3 \textwidth]{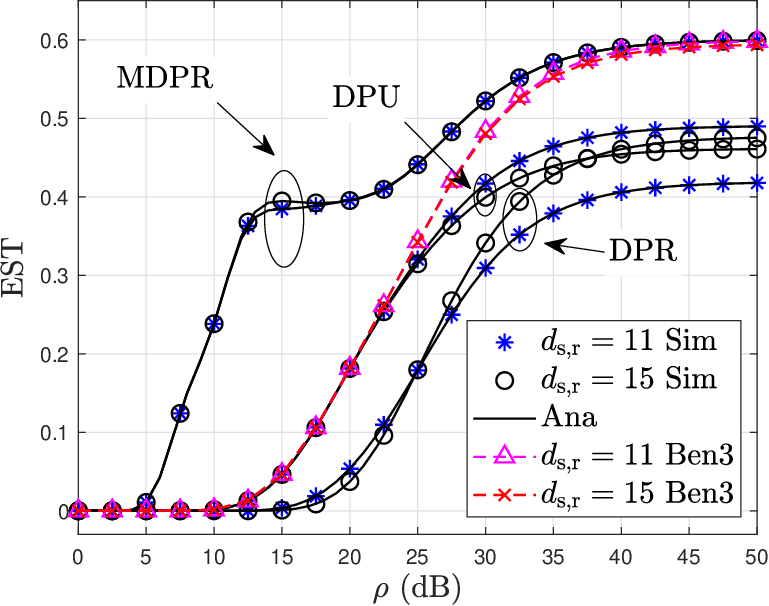}}
	\caption{OPs and EST for varying ${\rho}$ and ${d_{{\rm{s}},{\rm{r}}}}$.}
	\label{fig3}
\end{figure}
Fig. \ref{fig3} demonstrates the OPs and EST vs $d_{\rm{s,r}}$ for varying ${\rho}$ and ${d_{{\rm{s}},{\rm{r}}}}$.
From Fig.  \ref{fig31}, it can be observed that, as $d_{\rm{s,r}}$ increasing, the OP of $x_1$ for the DPR scheme decreases to a constant.
The reason is,
as $d_{\rm{s,r}}$ increases, $a_1$ decreases ($a_2$ increases), then the achievable rate at $U_1$ decoding $x_2$ increases, then $P_{{\rm{out}}}^{{x_1}}$ eventually converges to a constant, which equals to the ratio of $a_2$ to $a_1$.
Moreover, increasing $d_{\rm{s,r}}$, $P_{{\rm{out}}}^{{x_1}}$ of the MDPR scheme decreases and then increases, which is due to the fact that the OMA and NOMA scheme dominates in the lower-$\rho$ and the larger-$\rho$ region, respectively, and $P_{{\rm{out}}}^{{x_1}}$ of the OMA and NOMA scheme is decreasing/increasing with increasing $d_{\rm{s,r}}$.
$P_{{\rm{out}}}^{{x_1}}$ of the DPU scheme is independent of $d_{\rm{s,r}}$ because the power allocation coefficient is independent of $d_{\rm{s,r}}$.
In Fig. \ref{fig32}, $P_{{\rm{out}}}^{{x_2}}$ for the DPU and DPR schemes with larger $d_{\rm{s,r}}$ underperfoms that with lower $d_{\rm{s,r}}$ because of the increased path loss. 
The effect of $d_{\rm{s,r}}$ on $P_{{\rm{out}}}^{{x_2}}$ with the MDPR scheme is almost negligible since decoding $x_2$ is guaranteed at $R$ in MDPR scheme and $P_{{\rm{out}}}^{{x_2}}$ mainly depends on the quality of the second hop.
It is worth noting in Fig.  \ref{fig33} $d_{\rm{s,r}}$ has an different effect on $P_{{\rm{out}}}^{{x_2}}$ with the DPR scheme in the lower-$\rho$ and the large-$\rho$ region.
Further, $d_{\rm{s,r}}$ does not affect $P_{{\rm{out}}}^{{x_2}}$ with the MDPR and DPU schemes. 
This is because $x_2$ is successfully decoded in the first time slot then the $x_2$ forwarded by R will not interfere with $x_3$.
In Fig. \ref{fig34}, one can observe $d_{\rm{s,r}}$ has no significant effect on the EST of the MDPR scheme.
This is because $d_{\rm{s,r}}$ only affects $P_{{\rm{out}}}^{{x_2}}$ with the MDPR.
In the lower-$\rho$ region, the EST with DPU scheme outperforms that with the DPR scheme, which verifies the necessity to ensure that $U_1$ can successfully decode $x_2$.
	
\begin{figure}[!t]
	\centering
	\subfigure[OP of $x_1$]{
		\label{fig41}
		\includegraphics[width = 0.3 \textwidth]{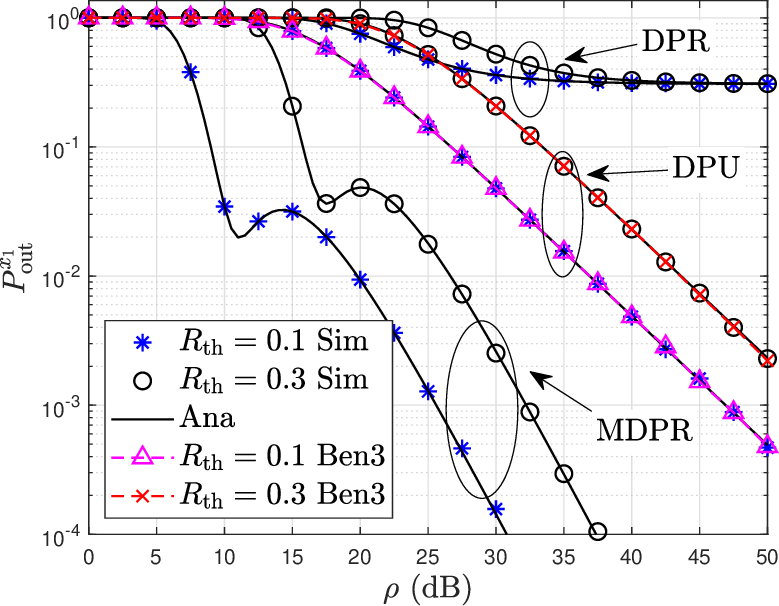}}
	\subfigure[OP of $x_2$]{
		\label{fig42}
		\includegraphics[width = 0.3 \textwidth]{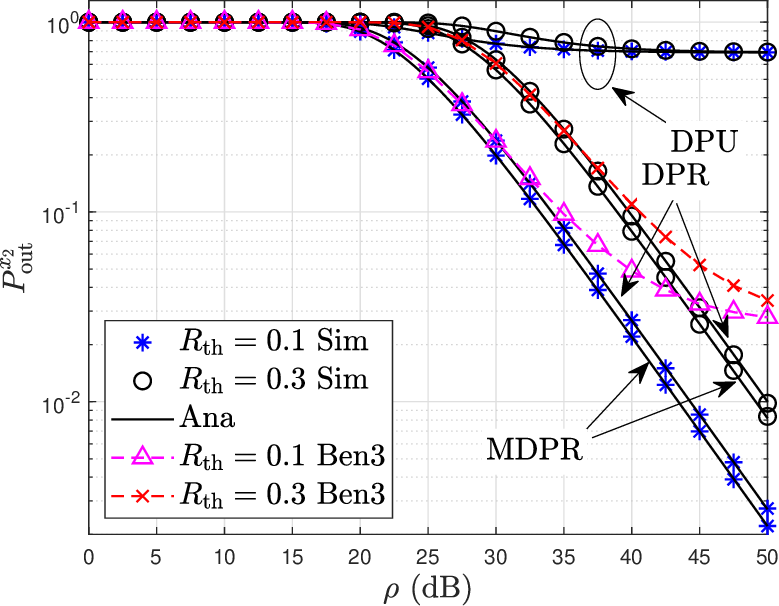}}
	\subfigure[OP of $x_3$]{
		\label{fig43}
		\includegraphics[width = 0.3 \textwidth]{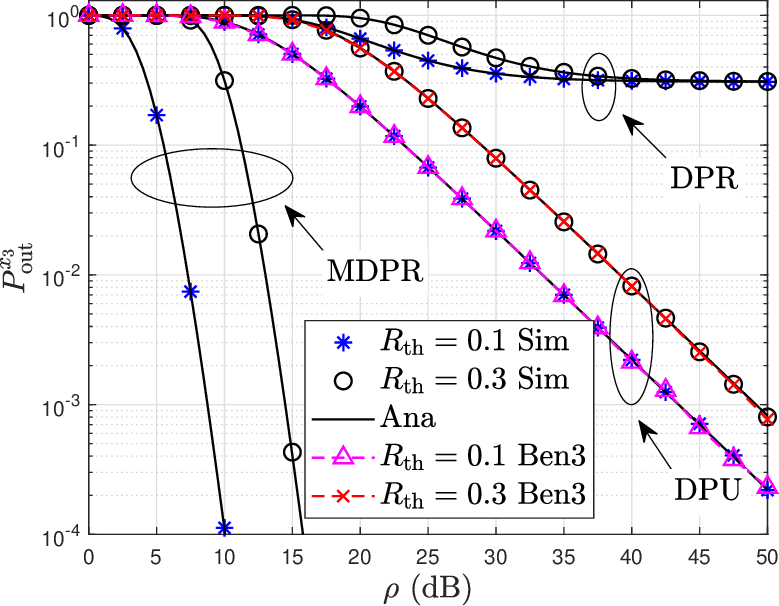}}
	\subfigure[EST]{
		\label{fig44}
		\includegraphics[width = 0.3 \textwidth]{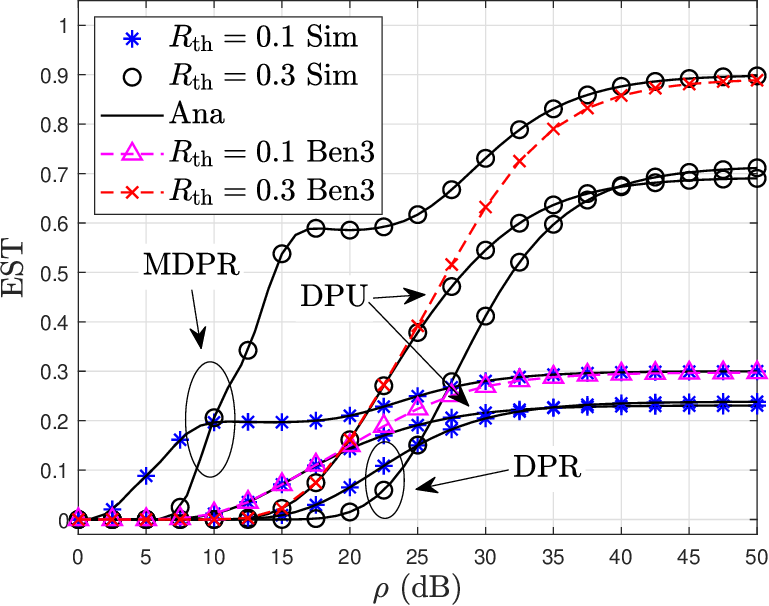}}
	\caption{OPs and EST for varying ${\rho}$ and $R_{\rm{th}}$.}
	\label{fig4}
\end{figure}
Fig. \ref{fig4} plots the effect of $\rho$ and $R_{\rm{th}}$ on the OPs and EST.
From Figs. \ref{fig41} - \ref{fig43}, it is easy to observe that the outage performance of $x_i$ worsens due to the higher rate requirement caused by increasing $R_{\rm{th}}$.
In Fig.  \ref{fig44}, one can observe that EST with lower $R_{\rm{th}}$ outperforms that with larger $R_{\rm{th}}$ in the lower-power region because the OP of $x_1$ and $x_3$ dominates the EST.
Moreover, the EST for the DPU scheme outperforms that for the DPR in the lower-power region because the outage performance of $x_1$ with the DPU scheme is superior to that with the DPR scheme.


\begin{figure}[!t]
	\centering
	\subfigure[OP of $x_1$]{
		\label{fig61}
		\includegraphics[width = 0.3 \textwidth]{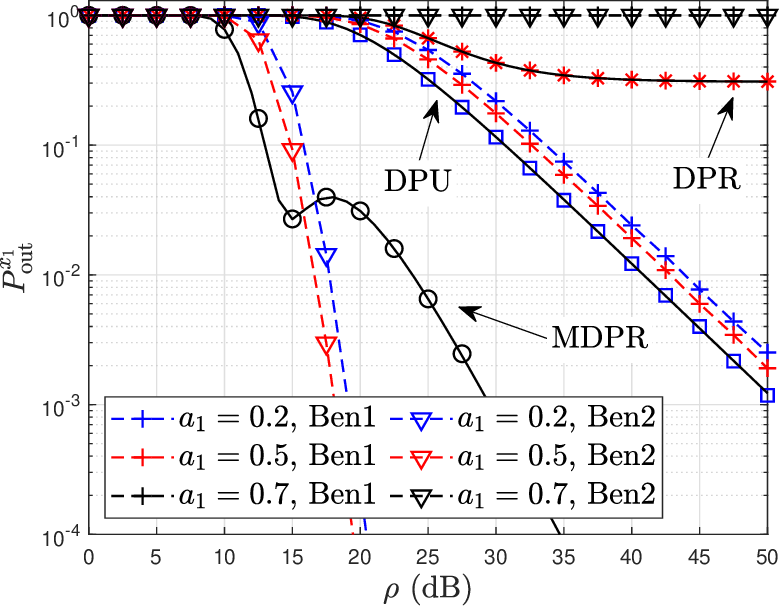}}
	\subfigure[OP of $x_2$]{
		\label{fig62}
		\includegraphics[width = 0.3 \textwidth]{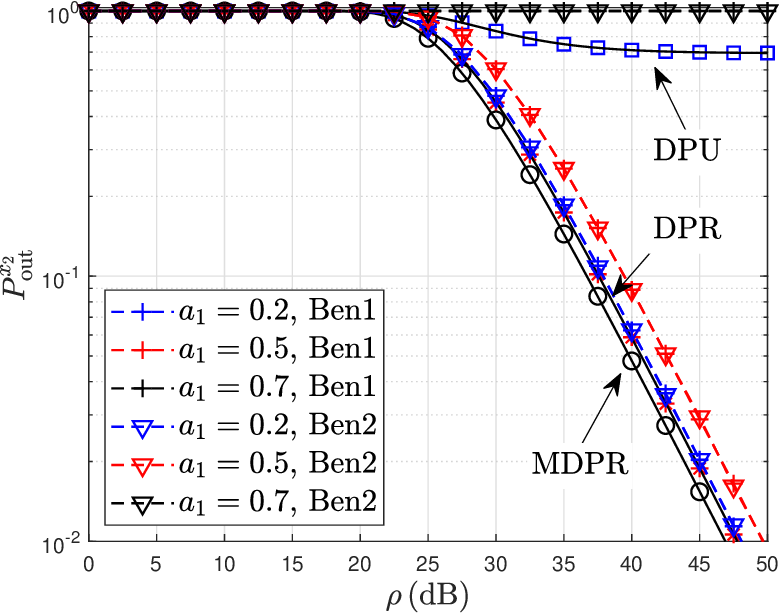}}
	\subfigure[OP of $x_3$]{
		\label{fig63}
		\includegraphics[width = 0.3 \textwidth]{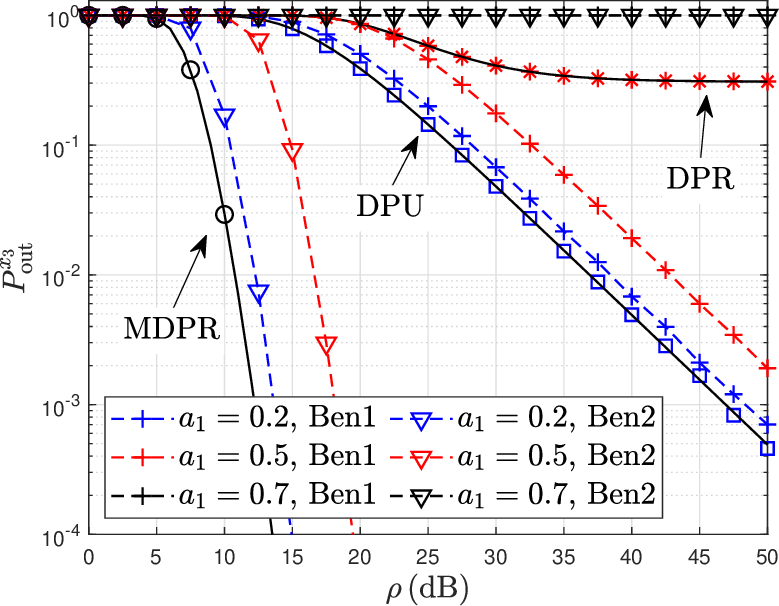}}
	\subfigure[EST]{
		\label{fig64}
		\includegraphics[width = 0.3 \textwidth]{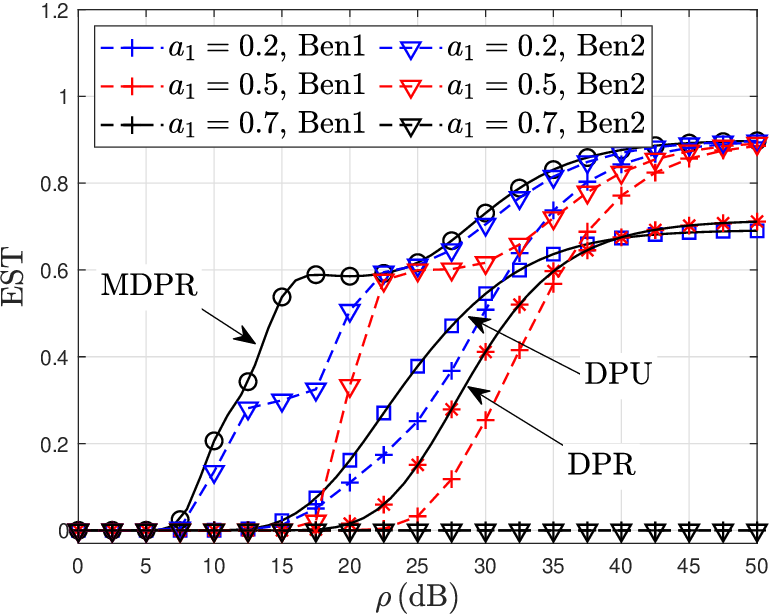}}	
	\caption{OPs and EST for varying $a_1$ and ${\rho}$.}
	\label{fig6}
\end{figure}
Fig. \ref{fig6} demonstrates the OPs and EST vs $\rho$ for varying $a_1$.
One can observe from Fig. \ref{fig61} that $P_{{\rm{out}}}^{{x_1}}$ of Ben1 and Ben2 with $a_1 = 0.2$ worse than that with $a_1 = 0.5$, but better than that with $a_1 = 0.7$, which indicating the existence of an optimal power allocation factor and reflecting the advantages of DPA scheme.
Moreover, $P_{{\rm{out}}}^{{x_1}}$ with the DPR scheme is worse than that with the DPU scheme because $U_1$ can decode $x_1$ with the condition that $x_2$ can be decoded successfully.
Fig. \ref{fig62} shows that $P_{{\rm{out}}}^{{x_2}}$ with the DPR scheme is superior to that with the DPU scheme because the DPR scheme improves the performance of $x_2$ and the DPU scheme is designed to enhance the performance of $x_1$.
Fig. \ref{fig63} demonstrates that $P_{{\rm{out}}}^{{x_3}}$ with the DPU scheme is preferred to that with the DPR scheme because parallel transmission can be realized only when $U_1$ can decode $x_2$.
Based on Figs. \ref{fig24}, \ref{fig34}, \ref{fig44}, and \ref{fig64}, one can find that the proposed MDPR scheme can improve the performance of the NOMA-based CDRT systems in the lower-$\rho$ region.


\begin{figure}[!t]
	\centering
	\includegraphics[width = 2.5 in]{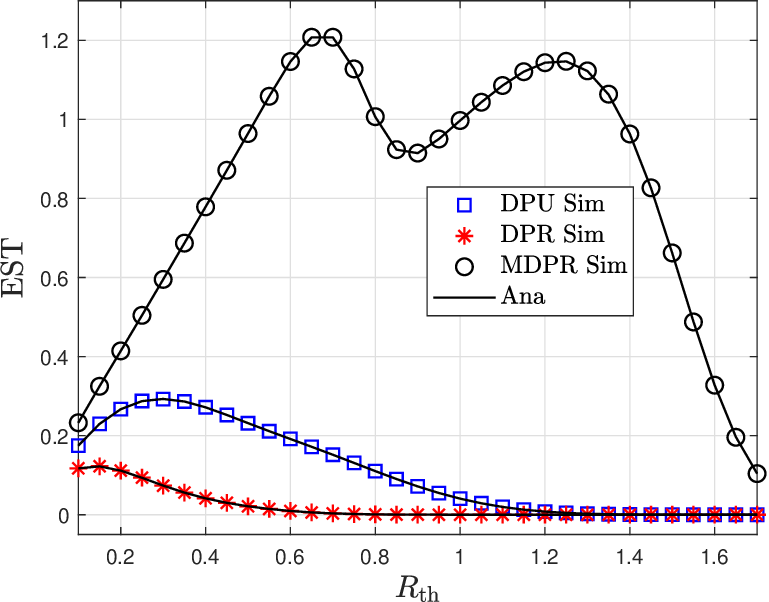}
	\caption{EST vs $R_{\rm{th}}$ with $\rho = 23$ dB.}
	\label{fig5}
\end{figure}
Fig. \ref{fig5} provides a comparison of the EST versus $R_{\rm{th}}$ with $\rho = 23$ dB.
It is easy to observe that there is an optimal $R_{\rm{th}}$ to maximize the EST for both the DPR and DPU scheme because $R_{\rm{th}}$ increases faster than that of the OP for the signals, results in the increasing of the EST. As $R_{\rm{th}}$ increases, the OP deteriorates; thus, the EST decreases.
For the MDPR scheme, there are two $R_{\rm{th}}$ to optimize the EST of the considered systems, corresponding to $x_1$ and $x_3$, respectively.


\begin{figure}[!t]
	\centering
	\includegraphics[width = 2.5 in]{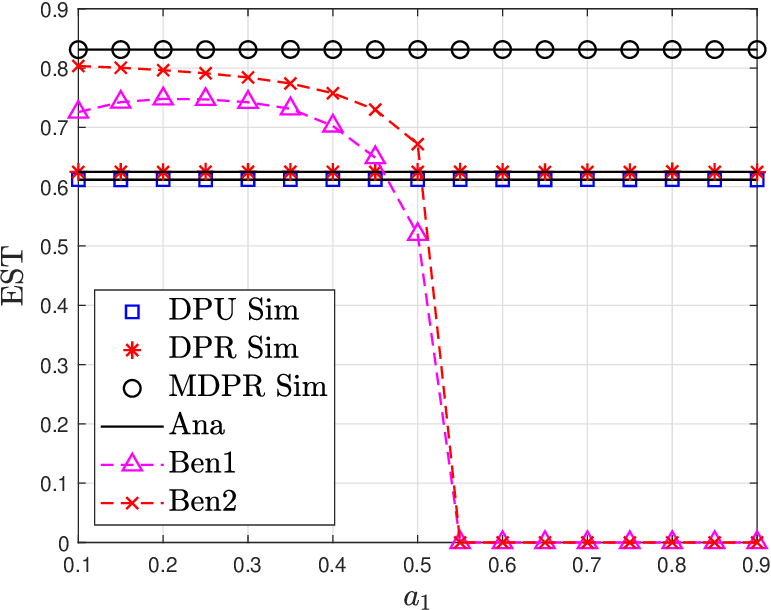}
	\caption{EST for varying $a_1$ with $\rho = 35$ dB.}
	\label{fig7}
\end{figure}
Fig. \ref{fig7} presents the impact of $a_1$ on the EST with $\rho = 35$ dB.
One can observe that the EST of Ben1 initially increases and subsequently decreases, and that of Ben2 decreases with increasing $a_1$.
The OPs of the two benchmark approach 1 when $a_1$ exceeds 0.55, then the EST is approximately equal to 0.
This is because, as $a_1$ increases, $x_2$ can not be decoded at $U_1$ in the first time slot and parallel transmission cannot be realized in the second time slot. This verifies that the DPA scheme outperforms that with the FPA scheme.

\section{Conclusion}
\label{sec: Conclusion}
In this paper, we proposed an adaptive scheme to provide reliability for the CDRT system through DPA and beamforming strategy. To ensure the CDRT system is implemented correctly, we designed the power allocation scheme to guarantee the relay can successfully decode the message for the edge-user with the DPA scheme. The beamforming scheme was utilized to ensure the center-user can remove the user interference and achieve parallel transmission.
To characterize the reliable performance of the proposed schemes, we derived the expression for exact OP and EST and analyzed the effects of system parameters on the EST in detail. The correctness of the analytical results was verified through Monte Carlo simulation. Simulation results demonstrated that the proposed adaptive CDRT scheme eliminated error floor for the edge-user and achieved better reliability than the benchmark scheme. In future work, our focus will seamlessly transition into unraveling the enigmatic layers of secrecy performance within the adaptive CDRT paradigm.



\begin{thebibliography}{1}

\bibitem{DingZ2017JSAC}
Z. Ding, X. Lei, G. K. Karagiannidis, and R. Schober,  ``A survey on non-orthogonal multiple access for 5G networks: Research challenges and future trends," \emph{IEEE J. Sel. Areas Commun.}, vol. 35, no. 10, pp. 2181-2195, Oct. 2017.

\bibitem{LvL2018Mag}
L. Lv, J. Chen, Q. Ni, Z. Ding, and H. Jiang, ``Cognitive non-orthogonal multiple access with cooperative relaying: A new wireless frontier for 5G spectrum sharing," IEEE Commun. Mag., vol. 56, no. 4, pp. 188-195, Apr. 2018.

\bibitem{WanD2018WC}
D. Wan, M. Wen, F. Ji, H. Yu, and F. Chen, ``Non-orthogonal multiple access for cooperative communications: Challenges, opportunities, and trends," IEEE Wireless Commun., vol. 25, no. 2, pp. 109-117, Apr. 2018.

\bibitem{ZengM2020Net}
M. Zeng, W. Hao, O. A. Dobre, and Z. Ding, ``Cooperative NOMA: State of the art, key techniques, and open challenges," IEEE Netw., vol. 34, no. 5, pp. 205-211, Sept. 2020.

\bibitem{KimJ2015CL}
J.-B. Kim and I.-H. Lee, ``Capacity analysis of cooperative relaying systems using non-orthogonal multiple access,'' \emph{IEEE Commun. Lett.}, vol. 19, no. 11, pp. 1949-1952, Nov. 2015.

\bibitem{LuoS2017CL}
S. Luo and K. C. Teh, ``Adaptive transmission for cooperative NOMA system with buffer-aided relaying,'' \emph{IEEE Commun. Lett.}, vol. 21, no. 4, pp. 937-940, Apr. 2017.

\bibitem{MukherjeeA2022TWC}
A. Mukherjee, P. Chakraborty, S. Prakriya, and A. K. Mal, ``Performance of novel adaptive schemes for cognitive full-duplex relaying based downlink cooperative NOMA,'' \emph{IEEE Trans. Wireless Commun.}, vol. 22, no. 5, pp. 3161-3179, Oct. 2023.

\bibitem{LiuYW2018WC}
Y. Liu, H. Xing, C. Pan, A. Nallanathan, M. Elkashlan, and L. Hanzo, ``Multiple antenna assisted non-orthogonal multiple access," \textit{IEEE Wireless Commun.}, vol. 25, no. 2, pp. 17-23, Apr. 2018.

\bibitem{MenJ2015CL}
J. Men and J. Ge, ``Non-orthogonal multiple access for multiple-antenna relaying networks,'' \emph{IEEE Commun. Lett.}, vol. 19, no. 10, pp. 1686-1689, Oct. 2015.

\bibitem{HanL2021CL}
L. Han, W.-P. Zhu, and M. Lin, ``Outage analysis of NOMA-based multiple-antenna hybrid satellite-terrestrial relay networks," \emph{IEEE Commun. Lett.}, vol. 25, no. 4, pp. 1109-1113, Apr. 2021.

\bibitem{LvL2020TWC}
L. Lv, Q. Ye, Z. Ding, Z. Li, N. Al-Dhahir, and J. Chen, ``Multi-antenna two-way relay based cooperative NOMA," \emph{IEEE Trans. Wireless Commun.}, vol. 19, no. 10, pp. 6486-6503, Oct. 2020.

%

\bibitem{KimJB2015CL}
J.-B. Kim and I.-H. Lee, ``Non-orthogonal multiple access in coordinated direct and relay transmission," \textit{IEEE Commun. Lett.}, vol. 19, no. 11, pp. 2037-2040, Nov. 2015.

\bibitem{LiX2020CL}
X. Li, Y. Chen, P. Xue, G. Lv, and M. Shu, ``Outage performance for satellite-assisted cooperative NOMA systems with coordinated direct and relay transmission," \emph{IEEE Commun. Lett.}, vol. 24, no. 10, pp. 2285-2289, Oct. 2020.

\bibitem{ZouL2020CL}
L. Zou, J. Chen, L. Lv, and B. He, ``Capacity enhancement of D2D aided coordinated direct and relay transmission using NOMA," \emph{IEEE Commun. Lett.}, vol. 24, no. 10, pp. 2128-2132, Oct. 2020.

%
%
%

\bibitem{XuY2018IET}
Y. Xu, G. Wang, L. Zheng, and S. Jia, ``Performance of NOMA-based coordinated direct and relay transmission using dynamic scheme," \textit{IET Commun.}, vol. 12, no. 18, pp. 2231-2242, Oct. 2018.

\bibitem{XuY2021TWC}
Y. Xu, J. Cheng, G. Wang, and V. C. M. Leung, ``Adaptive coordinated direct and relay transmission for NOMA networks: A joint downlink-uplink scheme," \emph{IEEE Trans. Wireless Commun.}, vol. 20, no. 7, pp. 4328-4346, Jul. 2021.

\bibitem{YuanL2022WCL}
L. Yuan, N. Yang, F. Fang, Q. Du, and Z. Zheng, ``Optimal power allocation for finite blocklength cooperative NOMA with coordinated direct and relay transmission," \textit{IEEE Wireless Commun. Lett.}, vol. 11, no. 3, pp. 523-527, March. 2022.

\bibitem{AnandJ2022TCOM}
A. Jee, K. Agrawal, and S. Prakriya, ''A coordinated direct AF/DF relay-aided NOMA framework for low outage,'' \emph{IEEE Trans. Commun.}, vol. 70, no. 3, pp. 1559-1579, Mar. 2022.


\bibitem{AnandJ2022IoT}
A. Jee and S. Prakriya, ``Performance of energy and spectrally efficient AF relay-aided incremental CDRT NOMA based IoT network with imperfect SIC for smart cities,'' \emph{IEEE Internet Things J.}, doi: 10.1109/jiot.2022.3229102, pp. 1-1,  2022.




\bibitem{LeiH2023CDRT}
H. Lei, X. She, K.-H. Park, I. S. Ansari, Z. Shi, J. Jiang, and M.-S. Alouini, ``On secure NOMA-CDRT systems with physical layer network coding," \textit{IEEE Trans. Commun.}, vol. 71, no. 1, pp. 381-396, Jan. 2023.

\bibitem{VuT2022TVT}
T.-H. Vu, T.-V. Nguyen, Q.-V. Pham, D. B. d. Costa, and S. Kim, ``Hybrid long-and short-packet based NOMA systems with joint power allocation and beamforming design,'' \emph{IEEE Trans. Veh. Technol.}, vol. 72, no. 3, pp. 4079-4084, Mar. 2022.

\bibitem{LeiH2022IoT}
H. Lei, C. Zhu, K.-H. Park, W. Lei, I. S. Ansari, and T. A. Tsiftsis, ``On secure NOMA-based terrestrial and aerial IoT systems,'' \emph{IEEE Internet Things J.}, vol. 9, no. 7, pp. 5329-5343, Apr. 2022.


\bibitem{LeiH2017TGCN}
H. Lei, M. Xu, I. S. Ansari, G. Pan, K. A. Qaraqe, and M.-S. Alouini, ``On secure underlay MIMO cognitive radio networks with energy harvesting and transmit antenna selection," \textit{IEEE Trans. Green Commun. Netw.}, vol. 1, no. 2, pp. 192-203, Jun. 2017.


\bibitem{ZhongC2015TCOM}
C. Zhong, X. Chen, Z. Zhang, and G. K. Karagiannidis, ``Wireless-powered communications: Performance analysis and optimization,'' \textit{IEEE Trans. Commun.}, vol. 63, no. 12, pp. 5178-5190, Dec. 2015.



\bibitem{YanM2012WCL}
M. Yan, Q. Chen, X. Lei, T. Q. Duong, and P. Fan, ``Outage probability of switch and stay combining in two-way amplify-and-forward relay networks," \emph{IEEE Wireless Commun. Lett.}, vol. 1, no. 4, pp. 296-299, Aug. 2012.

\bibitem{Gradshteyn2007Book}
I. S. Gradshteyn and I. M. Ryzhik, \emph{Table of Integrals, Series, and Products}, 7th edition. Academic Press, 2007.

%
%
%

\end{thebibliography}
\end{document}